\documentclass[fleqn,usenatbib]{mnras}

\usepackage{newtxtext,newtxmath}
\usepackage[T1]{fontenc}

\DeclareRobustCommand{\VAN}[3]{#2}
\let\VANthebibliography\thebibliography
\def\thebibliography{\DeclareRobustCommand{\VAN}[3]{##3}\VANthebibliography}

\usepackage{graphicx}	% Including figure files
\usepackage{amsmath}	% Advanced maths commands

\newcommand{\halfspace}{\hspace{1pt}}

\newcommand{\Lya}{Ly$\alpha$}
\newcommand{\Lyb}{Ly$\beta$}
\newcommand\HI{{\hbox{H\halfspace$\rm \scriptstyle I$}}}
\newcommand\HII{{\hbox{H\halfspace$\rm \scriptstyle II$}}}
\newcommand\HeI{{\hbox{He\halfspace$\rm \scriptstyle I$}}}
\newcommand\HeII{{\hbox{He\halfspace$\rm \scriptstyle II$}}}
\newcommand\HeIII{{\hbox{He\halfspace$\rm \scriptstyle III$}}}

\newcommand\lsim{~\lower.5ex\hbox{$\buildrel < \over \sim$}~}
\newcommand\gsim{~\lower.5ex\hbox{$\buildrel > \over \sim$}~}

\title[The 21-cm signature of galactic X-ray heating]{The 21-cm signature of X-ray heated halos around galaxies during cosmic dawn}

\author[Leong and Meiksin]{
Ka-Hou Leong,$^{1}$
A. Meiksin,$^{1}$\thanks{E-mail: aam@roe.ac.uk (AAM)}
\\
$^{1}$Institute for Astronomy, University of Edinburgh, Blackford Hill, Edinburgh EH9 3HJ UK
}

\date{Accepted XXX. Received YYY; in original form ZZZ}

\pubyear{\the\year{}}

\begin{document}
\label{firstpage}
\pagerange{\pageref{firstpage}--\pageref{lastpage}}
\maketitle

% Abstract of the paper
\begin{abstract}
X-rays emitted by high mass X-ray binaries (HMXBs) and supernovae-driven winds in the first galaxies during Cosmic Dawn are expected to warm the intergalactic medium prior to its reionization. While most of the heating will be uniform on measurable scales, exceptionally bright sources will produce a warm ring around them with a distinctive 21-cm signature. The detection of such systems would confirm X-rays are a source of IGM heating during Cosmic Dawn and provide a test of models predicting higher X-ray luminosities per star formation rate compared with present-day galaxies. We illustrate the effect for a star-forming galaxy in a $10^{11}\, M_\odot$ halo at $z=12$, treating the photoionizing radiation and X-rays using a novel fully time-dependent 3D ray-tracing radiative transfer code. We consider a range in possible spectra for the HMXBs and star formation efficiencies, as well as the possible effect of an extended halo around the galaxy. We find detection of the signal would require integration times of a few thousand hours using SKA1-Low except for a bright galaxy like a starburst, but only a thousand hours for the expected noise levels of SKA2-Low. Depending on the surrounding gas density profile, the 21-cm signature of X-ray heating may still require an exceptionally high star formation rate, either intrinsic to the source or provided by other systems clustered near it, to avoid dominance of the signal by absorption from the surrounding gas.
\end{abstract}

% Select between one and six entries from the list of approved keywords.
% Don't make up new ones.

%%%%%%%%%%%%%%%%%%%%%%%%%%%%%%%%%%%%%%%%%%%%%%%%%%

%%%%%%%%%%%%%%%%% BODY OF PAPER %%%%%%%%%%%%%%%%%%

\section{Introduction}

The first galaxies arising at Cosmic Dawn are expected to become detectable through their impact on the still neutral intergalactic medium (IGM). Continuum radiation redshifted to the local \Lya\ frequency will decouple the hyperfine spin state of hydrogen from the Cosmic Microwave Background (CMB), with the spin temperature taking on a value intermediate between the CMB and IGM temperatures \citep{1997ApJ...475..429M} through the Wouthuysen-Field effect \citep{1952AJ.....57R..31W, 1958PIRE...46..240F}. Either an aborption or emission signature against the CMB will result, depending on whether the IGM gas temperature is below or above the CMB temperature, respectively.

Initially the IGM temperature will be below the CMB temperature as a result of adiabatic expansion following the recombination era. Subsequent X-ray emission from galaxies, however, may increase the IGM temperature to above that of the CMB before the IGM is reionized. The most likely sources are X-ray binaries, especially high mass X-ray binaries (HMXBs) \citep{1997ApJ...475..429M, 2013ApJ...776L..31F, 2014Natur.506..197F, 2017ApJ...840...39M},
and supernovae-driven winds \citep{2017MNRAS.471.3632M}, suggesting the IGM will be heated to a temperature above the CMB temperature at some time in the redshift range $7<z<13$.

The X-ray luminosity from HMXBs in present-day galaxies is proportional to the star formation rate \citep{2012MNRAS.419.2095M}. Because of the expected low metallicity of the stars in high redshift galaxies, more massive accreting black holes are predicted, resulting in more luminous HMXBs by an order of magnitude at 2--10~keV \citep{2017ApJ...840...39M}. If dominated by Pop~III stars, the X-ray luminosity per star formation rate may be as much as a factor 30 brighter than current HMXBs \citep{2023MNRAS.521.4039S}.
 
By the time X-ray heating becomes important ($z<15$), the number of galaxies is sufficiently large that X-ray heating is expected to be homogeneous \citep{2014Natur.506..197F, 2017ApJ...840...39M}. Proximity heating zones, where the heating of a single galactic source dominates the background, will typically be small compared with the sub-Mpc mean spacing between sources \citep{2017ApJ...840...39M}, and consequently unresolvable by upcoming 21-cm experiments, precluding direct confirmation that X-rays emanating from typical galaxies are the source of the heating. In this paper, we examine the circumstances under which rare, unusually bright systems may be distinguished from the general population, producing a region of high IGM temperature relative to the mean temperature outside of any photoionized region. Such a region will produce an extended image of reduced 21-cm absorption against the CMB compared to the diffuse surroundings. The detection of such systems in 21-cm imaging experiments would provide direct evidence for X-ray heating of the IGM by at least some galaxies. Comparison with an independent estimate of the star formation rate in the galaxies would also provide a test of the expected high ratio of X-ray luminosity to star formation rate compared with present-day galaxies.

To perform the computations, we introduce a three-dimensional time-dependent adaptive ray-tracing radiative transfer method, for which the radiation propagates at the speed of light. Since the IGM is optically thin to the X-rays that warm it, even during Cosmic Dawn when the IGM is still predominantly neutral, the range over which an individual source heats the IGM expands near the speed of light. For a very bright source, the local X-ray brightness exceeds the background in a large proximity zone of a few proper megaparsecs. The expanding heating front will distort the observed images because of the equal arrival time effect:\ the measured image corresponds to an extended region for which the signal arrives simultaneously at the telescope. 
Although instantaneous radiative transfer methods capture the equal arrival time effect along the line of sight (LOS) \citep{2007MNRAS.374..493B}, only a three-dimensional time-dependent method can account for the equal arrival time effect for the full image. The adaptive ray-tracing method \citep{2002MNRAS.330L..53A} moreover maintains the directionality of the X-ray heating zone in the far field and the shape of the X-ray heated region, in contrast to time-dependent diffusive moment-based methods \citep{2008MNRAS.387..295A, 2013ApJS..206...21S}, which may distort the region exposed to radiation \citep{2021JCAP...02..042W}.

In the next section, we describe the computation of the photoionisation and heating around a galaxy, followed by a Results section illustrating some scenarios and a Discussion section. We end with a Summary and Conclusions section. In the Appendix, we describe the time-dependent radiative transfer code used and two test problems, the role secondary electron ionisations play on the 21-cm signal and characterise the detectability of the X-ray heating signature by the first generation of the Square Kilometre Array (SKA). For the calculations, we assume a flat $\Lambda$CDM universe with the cosmological parameters $\Omega_m = 0.3111$, $\Omega_v = 1 - \Omega_m$, $\Omega h^2 = 0.02242$, $h = 0.6766$, $n = 0.9665$ and $\sigma_8 = 0.8102$ \citep{2018arXiv180706209P}. Comoving units are prefixed with a ``c'' and proper with a ``p''.

\section{Methods}
\label{sec:methods}
\subsection{Emission models}
\label{subsec:em_model}

We assume a two-component spectral model for the source luminosity arising from photoionizing stars and HMXBs. We model the photoionizing star component as a black-body spectrum with temperature $10^5$~K, appropriate for a galaxy spectrum dominated by Pop~III stars \citep{1984ApJ...280..825B}. The energy range used for the photon packages extends from $13.6\, \rm{eV}$ to $1\, \rm{keV}$. Observations and theoretical modelling of high latitude ionized gas around the Milky Way suggest that ionization radiation is transmitted largely perpendicular to the Galactic disk through channels opened by supernovae, with an escape fraction of $\sim10$ percent \citep{1989ApJ...345..372N, 1989ApJ...339L..29R, 1999ApJ...510L..33B}. Numerical hydrodynamical simulations of galaxies support approximately biconal emission of photoionizing radiation \citep{2003ApJ...599...50F, 2008ApJ...672..765G, 2016ApJ...833...84X}, while observations of high redshift galaxies are consistent with a small covering fraction of escaping photoionizing radiation \citep{2006ApJ...651..688S}. Accordingly, we assume the blackbody emission is biconal, positioned vertically in the sky in the figures presented, with a half opening angle of $26^\circ$, covering about $10$ percent of the sky. Both radiation sources are placed at the centre of the simulation volume, which contains $128^3$ cells and has a proper side length of $5\, \rm{pMpc}$.

The intrinsic spectra of HMXBs inferred from \emph{Chandra} observations in nearby galaxies in the $0.25-8$~keV band vary widely from source to source, with power-law spectra $L_E\sim E^{-\alpha_x}$ for hard ($\alpha_x=-0.49\pm0.10$), soft ($\alpha_x=1.37\pm0.16$) and super-soft ($\alpha_x=2.72\pm0.21$) sources \citep{2017MNRAS.468.2249S}. We assume a collective spectrum, $L_E=1.31\times10^{40}\left(E/\mathrm{keV}\right)^{-1.08}\left({\dot M_*}/{\mathrm{M_\odot\,yr^{-1}}}\right)\,\mathrm{erg\,s^{-1}\,keV^{-1}}$ for a star formation rate $\dot M_*$, for the intrinsic HMXB contribution, spanning the energy range from $54.4\, \rm{eV}$ to $10\, \rm{keV}$, and adopting the scaling found by \citet{2017MNRAS.468.2249S} for the average galactic collective spectrum from the HMXBs in the galaxy (containing a collection of super-soft, soft and hard sources), enhanced by a factor of ten to account for the expected low metallicity of high redshift sources \citep{2017ApJ...840...39M}. We note that during Cosmic Dawn, the collective HMXB spectrum from galaxies may differ from our assumed value. We also consider each spectral type separately.

For all the simulations, hydrogen and helium are initially neutral with a temperature of $12\, \rm{K}$, consistent with \citet{2017ApJ...840...39M}. The dark matter correlation length at $z=12$ is about $3.5h^{-1}\,\mathrm{ckpc}$, well below the few arcminute resolution of the SKA (1.0 arcmin corresponds to $2.0h^{-1}$~cMpc at $z=12$). We therefore assume a uniform density background, with a hydrogen number density of $4.5\times10^{-4}\, \rm{cm}^{-3}$, corresponding to the mean IGM density at redshift $z=12.3$ (matching the simulation output used). The mass-fraction abundances of hydrogen and helium are 0.76 and 0.24, respectively. Because the dark matter and gas density are expected to be enhanced near a high peak, we also consider some models with the galaxy placed at the centre of an extended density profile.

Absorption of the emitted X-ray spectrum arises mainly from neutral hydrogen and metals internal to the galaxy, and from hydrogen and helium in the surrounding IGM \citep{2017ApJ...840...39M}. For sub-solar metallicity and a typical internal neutral hydrogen column of around $10^{20}-10^{21}\,\mathrm{cm}^{-2}$ \citep[eg][]{2001ApJ...558...56H}, the emergent X-ray spectrum would be suppressed at energies below about $0.12-0.25$~keV. This is comparable to the amount of suppression from \HI\ and \HeII\ in the IGM \HII\ bubble produced by the photoionizing stars in the galaxy. Rather than modelling the internal absorption, uncertain especially at the redshifts of interest, for simplicity we assume that the low energy suppression is dominated by the \HII\ bubble around the galaxy. As a consequence, the evolution in the soft end of the spectrum impinging on the neutral IGM beyond will be governed by the evolution of the surrounding \HII\ bubble, with a cut-off in the spectrum typically at around 0.2--0.3~keV. Since these soft photons dominate the heating of the still neutral IGM near the galaxy, the heating rate and temperature of the IGM beyond the ionized bubble will evolve.

A galaxy of constant specific luminosity $L_\nu$ will heat neutral hydrogen in the surrounding diffuse IGM at position $\mathbf{r}$ from the galaxy at time $t$ at the rate per particle
\begin{equation}
G(\mathbf{r},t) = \int_{\nu_L}^\nu\,d\nu\,\frac{L_\nu}{4\pi r^2}\left(1-\frac{\nu_L}{\nu}\right)\sigma_\nu e^{-\tau_\nu(\mathbf{r},t)},
\label{eq:H}
\end{equation}
where $\nu_L$ is the frequency at the hydrogen Lyman edge, $r=\vert\mathbf{r}\vert$ and $\sigma_\nu$ is the photoionization cross section. A similar expression applies for neutral helium. We assume the galaxy can only negligibly photoionize singly-ionized helium. Because the galaxy will also photoionize the surrounding hydrogen and (neutral) helium, we have allowed for a time-dependent optical depth in direction $\mathbf{r}=r\mathbf{\hat n}$,
\begin{equation}
\tau_\nu(\mathbf{r},t) = \int_0^r\,dr^\prime\,n_\mathrm{HI}\left(r^\prime,t-\frac{\vert r-r^\prime\vert}{c}\right)\sigma_\nu,
\label{eq:taunu}
\end{equation}
with a similar expression for \HeI. The time-delay has been shown explicitly in the neutral hydrogen density as the \HI\ fraction evolves.

The heating rate by galactic HMXBs is expected to be extremely uniform. Allowing for star formation in halos above a threshold mass of atomically cooled haloes of $\sim1-2\times10^7\,h^{-1}\mathrm{M}_\odot$ \citep{1997ApJ...484..985H, 2011MNRAS.417.1480M}, the number of star-forming haloes will be about 3--5 per (cMpc)$^3$ at $z=12$, using the halo mass function of \citet{2007MNRAS.374....2R}. The mean free path of a 0.2~keV photon in the neutral IGM at $z=12$ is $\sim5$~cMpc, so that more than 1000 star-forming galaxies will contribute significantly to the heating of any point in the IGM. Direct \emph{JWST} observations show that the cosmic star formation rate is sufficient for strong Wouthuysen-Field coupling \citep{1952AJ.....57R..31W,1958PIRE...46..240F} of the 21-cm spin temperature to the gas kinetic temperature \citep{2023RNAAS...7...71M}. Since source continuum photons between the \Lya\ and \Lyb\ frequencies may travel up to a distance $(5/27)c/H(z)$, where $H(z)$ is the Hubble parameter, before redshifting into local \Lya\ photons, the \Lya\ scattering rate driving the Wouthuysen-Field mechanism will also be highly uniform.

The signal against the CMB will reflect fluctuations in the matter density field \citep{2000ApJ...528..597T}. In order to isolate the effects of source emissivity on the 21-cm signal, we assume the surrounding gas is uniform. This is a good approximation for most of the density field at high redshifts where the fluctuations on resolvable scales will be in linear. Since a rare source is expected to be located at the peak of an extended density profile, however, its surroundings will show an increased amount of matter clustering. We examine possible consequences of the enhanced clustering on the 21-cm signature in the Discussion section.

The ambient neutral IGM on large scales will be heated by X-rays from the global population of galaxies to a temperature of about 12~K at $z = 12$ \citep{2017ApJ...840...39M}. This will result in an absorption signal against the CMB with the observed brightness temperature differential, for 21-cm optical depth $\tau_{21}$ and spin temperature $T_S$,
\begin{eqnarray}
    \delta T &=& (1+z)^{-1}\left(T_S-T_\mathrm{CMB}\right)\left(1-e^{-\tau_{21}}\right)\nonumber\\
    &\simeq&(31\,\mathrm{mK})\left(\frac{1+z}{13}\right)^{1/2}\left(1-\frac{T_\mathrm{CMB}}{T_S}\right)
    \label{eq:dTIGM}
\end{eqnarray}
\citep[][eq.~(45)]{1997ApJ...475..429M},
giving $\delta T\simeq-61$~mK for CMB temperature $T_\mathrm{CMB}=2.725(1+z)$ and $T_S=12$~K, assuming a neutral hydrogen fraction near unity. The gas near a massive source, however, will be heated at an enhanced rate, giving rise to an extended 21-cm proximity effect:\ an extended region of relatively less absorption compared with the diffuse IGM. We solve for the evolution of the \HII\ bubble and the heating beyond using a novel time-dependent ray-tracing radiative transfer package, \texttt{3DPhRay}. A description of the code is provided in the Appendix.
%We have implemented the package in the gravity-hydrodynamics code \texttt{Enzo} \citep{2014ApJS..211...19B}.

\subsection{Simulated Equal Observing-Time Surface}
\label{subsec:equal_time_surface}
\begin{figure}
    \centering
    \includegraphics[width=0.7\columnwidth]{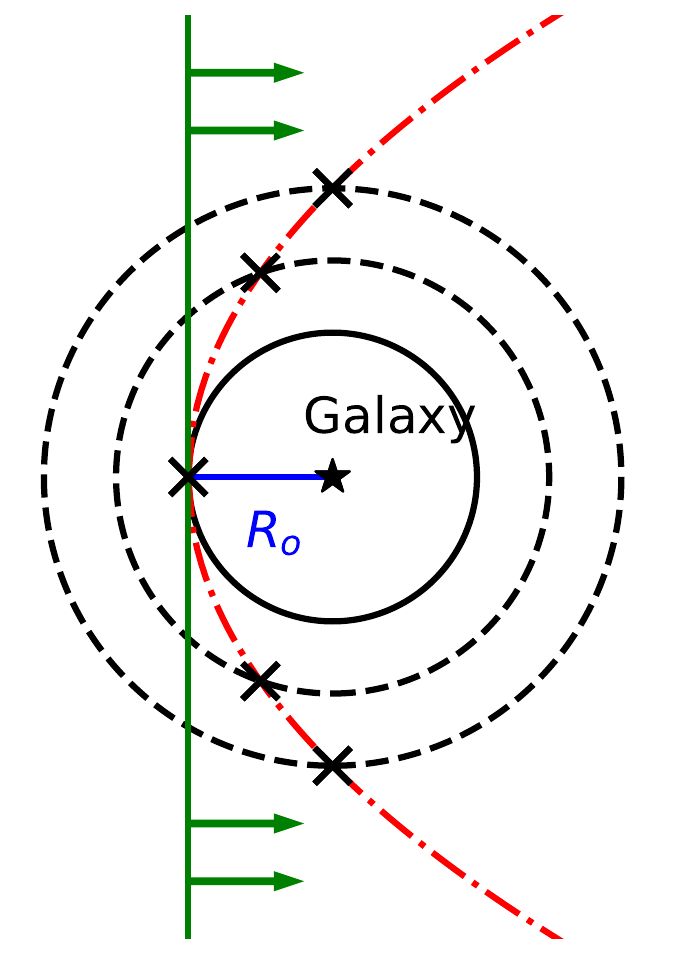}
    \caption{A 2-dimensional illustration of the observed region of the warm IGM bubble. The star marks the location of the galaxy. The circles represent the surfaces of the bubble at different moments, expanding at the speed of light.
    %%The blue solid line indicates the central axis along which the observer, the galaxy and the first touch point at $R_{0}$ are aligned.
    The green vertical line represents the EOTS on which the background CMB wavefront advances toward an observer on the right. The arrows indicate the direction of travel. The CMB front detected by the observer first contacts the bubble's surface at $R_{0} = c t_{0}$, indicated by the blue line, which is aligned with the galaxy and the observer. The red dash-dotted parabolic curve indicates the locations from which CMB light on the EOTS intersects the relativistically expanding bubble. This figure assumes the observer is sufficiently far away from the galaxy that the shape of the CMB wavefront is well approximated by a vertical surface, and that the observer's distance to the galaxy much exceeds the size of the warm bubble.
    }
    \label{fig:EOTS}
\end{figure}

During Cosmic Dawn, the warm IGM bubble generated by the X-ray emission for photons exceeding $\sim0.2$~keV from a first galaxy expands relativistically, as X-rays of these energies travel largely unabsorbed. This luminal expansion of the warm IGM bubble distorts the observed region of the bubble due to the effect of the equal observing-time surface (EOTS)\footnote{The equal observing-time surface is also known as the equal arrival time surface.}\citep{2005ApJ...634..715W}. The EOTS is a physical surface where the distances between the observer and each point on the surface are equal, meaning that the light from different points on the same EOTS arrives at the observer simultaneously. Given the cosmological distances of the first galaxies and that their warm IGM bubbles expand at the speed of light, the observed surface of the bubble takes the shape of a circular paraboloid. The surface may be expressed as
\begin{equation}
S^{2}[R_{0}, R(t)] \simeq 4R_{0}R(t) - 4 R_{0}^2,
\label{eq:L_EOTS}
\end{equation}
where $S[R_{0}, R(t)]$ is the distance between the parabolic surface and the central axis, $R(t) = ct$ denotes the radius of the bubble at time $t$ and $R_{0} = ct_{0}$ is the radius of the bubble at the moment when the EOTS first comes into contact with the bubble. This is shown in Fig.~\ref{fig:EOTS}, which illustrates the region of the warm IGM bubble the background CMB radiation passes through on its way to the observer.

Using the time-dependent ray-tracing method, it is straightforward to simulate the distortion caused by the effect of the EOTS. In \texttt{3DPhRay}, the propagation speed of the radiation field is set to the physical value of the speed of light, allowing us to simulate the radius of a luminally expanding bubble accurately. Photoionization heating and collisional and radiative recombination cooling are included. We include secondary electron ionizations as well since they reduce the heat deposited in the still neutral IGM by X-rays \citep{1997ApJ...475..429M}.

We use a post-processing procedure to reconstruct the observed region of the luminally expanding warm IGM halo. This is achieved by the following steps: firstly, we set the timestep in simulations to the travel time corresponding to a single cell, $\Delta t = \Delta l_{\rm {cell}} / c$, where $\Delta t$ is the length of the timestep and $\Delta l_{\rm {cell}}$ is the proper size of a simulation cell. Secondly, we configure \texttt{3DPhRay} to output a snapshot of the simulation at each timestep, so that the state of the simulation is stored for every timestep $\Delta t$. Finally, by combining slices from the related snapshots observed by the same EOTS which first touches the bubble at $R_{0}$, we reconstruct the observed region of the relativistic bubble (Fig.~\ref{fig:EOTS}). This construction ensures that every point in the post-processed data representing an observed image shares the same retarded time. We note that this procedure is well-suited for simulating the detected region of phenomena containing luminally expanding phases. For example, it may be applied to the ionisation processes of the IGM near a QSO that has recently turned on \citep{2023MNRAS.519.5743L}, for which the expansion velocity of the \HII\ ionisation front approaches the speed of light due to the high luminosity of the QSO, when ionizing photons are emitted at a rate too high for all to be absorbed within the light-front.

\subsection{Mock 21cm Signature Map on the Sky}
\label{subsec:mock_21cm_map}
After constructing the distorted warm IGM halo based on the effect of the EOTS, we compute the CMB spectrum on each line of sight (LOS). The neutral hydrogen gas along the LOS interacts with the CMB through the hyperfine structure of the hydrogen atom, absorbing and emitting the corresponding radio photons. We simulate the spectrum by solving the static space radiation transfer equation, which is expressed as:
\begin{equation}
  \frac{d I_{\nu}}{dl} = \frac{1}{c}\frac{\partial I_{\nu}}{\partial t} + \frac{\partial I_{\nu}}{\partial l} = - \kappa_{\nu} I_{\nu} + j_{\nu},
	\label{eq:static_RT}
\end{equation} 
where $I_{\nu} \equiv I(\nu,\mathbf{x},\hat{\Omega},t)$ is the specific intensity in units of energy per time per solid angle per unit area per frequency $\nu$, $\kappa_{\nu} \equiv \kappa (\nu,\mathbf{x},t)$ is the absorption coefficient and $j_{\nu} \equiv j (\nu,\mathbf{x},t)$ is the specific emissivity per solid angle. We neglect the energy dilution caused by the expansion of the Universe since the scale of the distinctive 21-cm signature created by a massive first galaxy is significantly smaller than the Hubble distance, $L_{\rm {bubble}} \ll c/H(z)$.

To solve Eq.~(\ref{eq:static_RT}), we trace $I_{\nu}$ through the LOS in the velocity coordinate. The observed $I_{\nu_j}$, where $\nu_j$ redshifts to the frequency of the 21-cm transition ($\nu_{10}$) at the $j^\mathrm{th}$ pixel on the LOS with Doppler velocity ${\bar v}_j$, is computed using a recurrence algorithm. At the start, $I_{\nu_j}({\bar v}_0)$, where ${\bar v}_0 = c (\lambda_\mathrm{21cm} - \lambda_0)/\lambda_0$ is the Doppler velocity at the zeroth pixel on the LOS with wavelength $\lambda_0$, is initialised to the black body spectrum, $I_{\nu_j}^\mathrm{BB}(T_{\mathrm{CMB}}[z])$, representing the CMB at redshift $z$. Then, $I_{\nu_j}({\bar v}_{i+1})$ is calculated by recursively applying the following formula:
\begin{equation}
  I_{\nu_j}(\bar {v}_{i+1}) = I_{\nu_j}(\bar {v}_{i})e^{-\kappa_{\nu_j}(\bar {v}_{i+1})\Delta x} + \frac{j_{\nu_j}(\bar {v}_{i+1})}{\kappa_{\nu_j}(\bar {v}_{i+1})}\left[1 - e^{-\kappa_{\nu_j}(\bar {v}_{i+1})\Delta x}\right],
	\label{eq:calculate_I}
\end{equation} 
where $\Delta x$ is the proper size of the pixel on the LOS. For the hyperfine structure of hydrogen, $\kappa_{\nu_j}$ and $j_{\nu_j}$ are given by
\begin{equation}
  \kappa_{\nu_j}(\bar {v}_{i}) = \frac{3h_{\rm P}\nu_{\rm {10}}}{c}\frac{T_\ast B_{\rm {10}}}{T_{S, i}}n^{0}_{\rm {HI, i}}\phi\left[\bar {v}_j - (\bar {v}_i + \bar {v}^\mathrm{pec}_i)\right],
	\label{eq:abs_kappa}
\end{equation} 
and
\begin{equation}
  j_{\nu_j}(\bar {v}_{i}) = h_{\rm P}\nu_{\rm {10}}A_{\rm {10}}n^{1}_{\rm {HI, i}}\phi\left[\bar {v}_j - (\bar {v}_i + \bar {v}^\mathrm{pec}_i)\right],
	\label{eq:emit_j}
\end{equation} 
respectively, where $n^{0}_{\rm {HI}}$ is the density of hydrogen atom in the singlet state, $n^{1}_{\rm {HI}}$ is the density of hydrogen atom in the triplet state, $T_\ast \equiv h_{\mathrm{P}}\nu_{10}/{k_\mathrm{B}}$, $A_{\rm{10}}=2.85\times10^{-15}\,\rm{s}^{-1}$ is the spontaneous decay rate, $B_{\rm {10}} = A_{\rm{10}}c^3/(8\pi h_{\rm P}\nu_{\rm {10}}^3)$ is the Einstein coefficient of stimulated emission, $\bar {v}^\mathrm{pec}_i$ is the peculiar velocity of the gas at the $i^\mathrm{th}$ pixel and $\phi(\bar {v})$ is the line profile in units of per frequency. The time delay effect is included by imposing the requirement that all the points in the reconstructed file have the identical retarded time. 

To account for the thermal broadening effect, we use a Gaussian profile as the line profile. Thus, the line profile for frequency $\nu_j$ at the $i^\mathrm{th}$ pixel is expressed as 
%$\phi(\bar {v}_j - (\bar {v}_i + \bar {v}^{pec}_i), \bar {v}_i^{\rm {th}})$:
\begin{equation}
  \phi\left[\bar {v}_j - (\bar {v}_i + \bar {v}^\mathrm{pec}_i), \bar {v}_i^{\mathrm{th}}\right] = \frac{c}{\nu_{\rm {10}}}\frac{1}{\bar {v}_i^{\rm {th}}\sqrt{2\pi}}\exp{\left\{-\frac{\left[\bar {v}_j - (\bar {v}_i + \bar {v}^\mathrm{pec}_i)\right]^2}{2(\bar {v}_i^{\mathrm{th}})^2}\right\}},
	\label{eq:line_profile}
\end{equation}
where $\bar {v}_i^{\rm {th}} = \sqrt{k_\mathrm{B}T_i/m_\mathrm{P}}$ is the thermal velocity. In practice, the Doppler effect contributed by the peculiar velocity of the gas can be ignored when the bandwidth is wide. For the 21-cm signature, considering that the bandwidth in a future observation is $1$ MHz, the equivalent velocity is about $2800\, \rm {km\,s^{-1}}$ at $z = 12$, which is significantly larger than peculiar velocities of the gas in the IGM. Accordingly, we take $\bar {v}^\mathrm{pec}=0$ and adopt a top-hat filter in the frequency direction in this work. 

The developing Square Kilometre Array (SKA) low-frequency component (SKA-low) is designed to measure the expected 21-cm signature during Cosmic Dawn \citep{2013ExA....36..235M}. Since the instrumental effects of the SKA-low have not yet been established, we use a simple routine to generate the mock radio map. We first take the average of $I_{\nu_j}$ on each LOS over the desired bandwidth. Then, we conduct a convolution in angle on the sky using a Gaussian kernel. Finally, we convert the smoothed $I_{\nu_j}$ map to a brightness temperature map and add Gaussian noise to the smoothed brightness temperature map.

\section{Results}
\label{sec:results}

\begin{figure*}
\scalebox{0.65}{\includegraphics{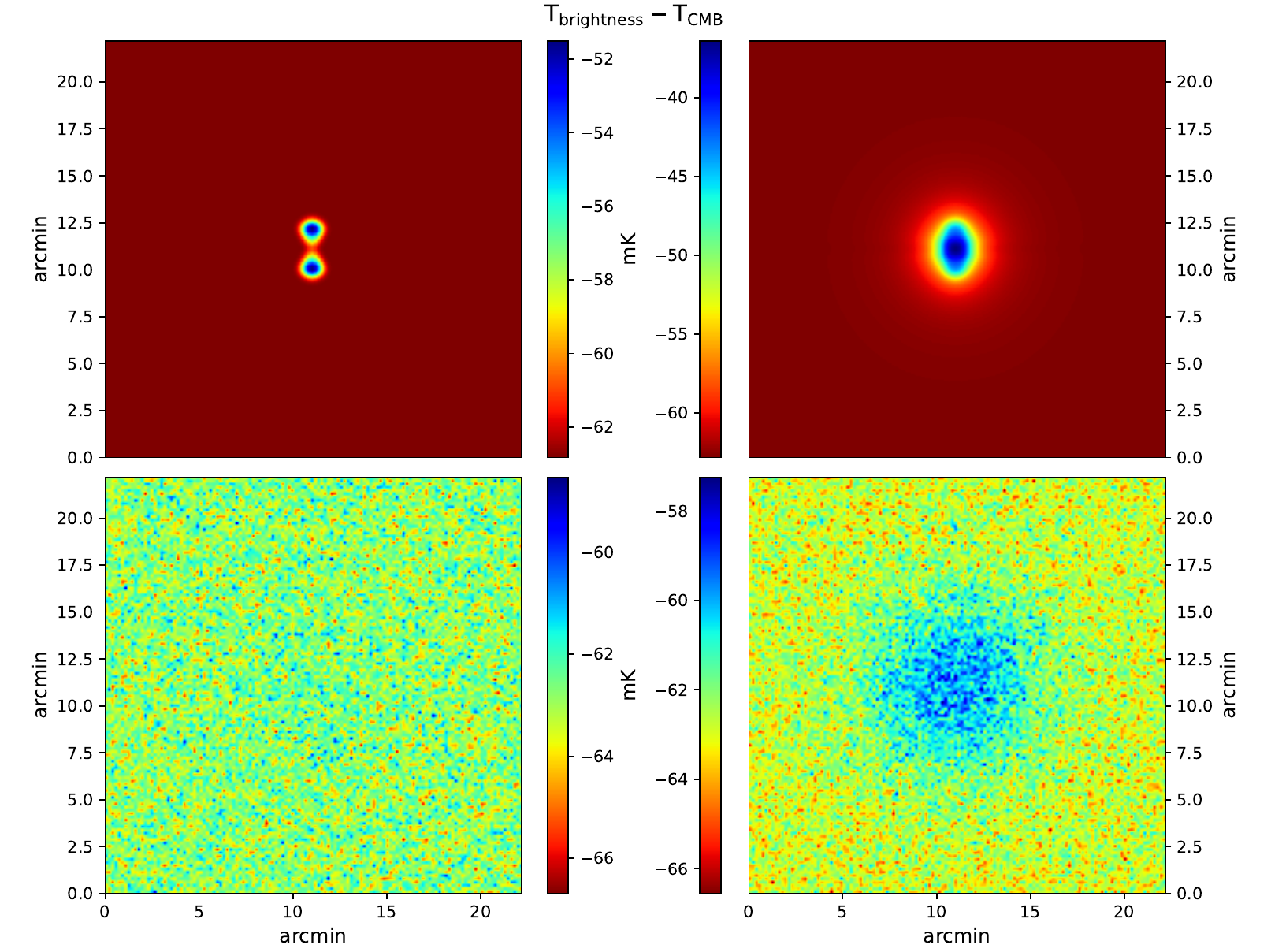}}
%\vspace{-1.5cm}
\caption{Top panels: The simulated temperature differential between the CMB and 21-cm IGM brightness as projected on the sky 7.8~Myr after the source turns on. The luminosity profile of the central galaxy shown in the left-hand panel contains emission only from its stellar component, represented as blackbody radiation. In addition to photons from the stellar component, the central galaxy in the right-hand panel includes X-ray emission contributed by HMXBs, modelled as a composite power-law spectrum $L_E\sim E^{-1.08}$ (see text). Both simulations adopt a SFR $13.8\,\mathrm{M_\odot\,yr^{-1}}$, and the low metallicity assumed at high redshift enhances the luminosity of the HMXBs compared with present-day galaxies by a factor of ten. The X-ray component from HMXBs heats the still neutral IGM surrounding the galaxy, weakening the IGM absorption signature in its vicinity. Bottom panels: The associated mock observed images. Both images have been Gaussian-smoothed with a FWHM of 7~arcmin, and Gaussian noise has been added with an rms of 1~mK, according to the expected detectability of SKA2-Low.
}
\label{fig:image_21cm_no_burst_BBR_-1.08_R_2.38}
\end{figure*}

To illustrate the local heating effect of a rare, massive galaxy, we consider a galaxy in a halo of mass $10^{11}\,\mathrm{M_\odot}$ at $z=12$, for which there will be on average 14 systems per (cGpc)$^3$. We estimate the star formation rate from the rate at which baryons collapse into halos near this mass \citep[eg][]{2005ApJ...626....1B}. Assuming a conversion efficiency of collapsed baryons into stars of one percent, the estimated star formation rate (SFR) is $13.8\,\mathrm{M_\odot\,yr^{-1}}$. For a conversion rate to hydrogen-ionizing photons of $2.5\times10^{53}\,\mathrm{ph\,s^{-1}\,M_\odot^{-1}\,yr}$, appropriate to a Salpeter initial mass function of Pop~III stars forming in the mass range 1--500~M$_\odot$ \citep{2010AA...523A..64R}, and allowing for an escape fraction of ionizing radiation from the galaxy of 0.3 \citep{2018MNRAS.478.4851I, 2023NatAs...7..611R}, we introduce a central photoionizing source of strength $S=3.5\times10^{54}\,\mathrm{ph\,s^{-1}}$. Only a fraction of this is radiated by a beamed source, reduced by the sky-covering fraction of the beam. Using the conversion $L_{1500} = 10^{40.4}\,\left({\dot  M_*}/{\mathrm{M_\odot\,yr^{-1}}}\right)\,\mathrm{erg\,s^{-1}\,A^{-1}}$ \citep{2010AA...523A..64R}, the galaxy would have magnitude $m_\mathrm{AB}=25.8$ in the \emph{JWST} F200W filter, well above the detection threshold of \emph{JWST} in a pointed observation \citep{2023NatAs...7..611R}. (The galaxy would be a magnitude dimmer allowing for stars to form down to 0.1~M$_\odot$; the same photoionizing strength would be preserved for an escape fraction of 0.7.) We show results about 8 million years (and later, allowing for time delay) after the source turns on, creating an ionization front about $0.21$~pMpc in radius, corresponding to $0.94$ arcmin on the sky. The 21-cm temperature differential $\delta T=T_B-T_\mathrm{CMB}$ is shown in Fig.~\ref{fig:image_21cm_no_burst_BBR_-1.08_R_2.38} allowing only for photoionization by the stellar component (left-hand panel), and after including the X-ray component from HMXBs as well (right-hand panel).

To model the detectability of the 21-cm signature, we smooth the image to represent the beam resolution and add the expected amount of noise for a SKA observation of 1000~hrs. For the results shown below, the assumed full-width half-maximum (FWHM) of the Gaussian point spread function (PSF) is $7$~arcmin, expected for the SKA in AA4 core only mode (SKAO Sensitivity Calculator at \texttt{https://sensitivity-calculator.skao.int}). For an effective collecting area $A_\mathrm{eff}$ and system temperature $T_\mathrm{sys}$, SKA1-Low is expected to achieve a sensitivity $A_\mathrm{eff}/ T_\mathrm{sys}\simeq1$~m$^2$~K$^{-1}$ per dish at 107~MHz \citep{2019arXiv191212699B, 2022PASA...39...15S} and SKA2-Low a factor 5--10 higher \citep{2019arXiv191212699B}. In $1000\, \rm{hrs}$ of integration time using a bandwidth (BW) of $1\, \rm{MHz}$, this corresponds to a Gaussian noise root mean square (rms) brightness temperature of 0.5--1~mK for SKA2-Low and 5~mK for SKA1-Low. We adopt a noise level of 1~mK for most of our models, but also show other cases appropriate to SKA1-Low in App.~\ref{appendix:signal_SKA1_low}. The results are computed by numerically integrating the radiative transfer equation for the 21-cm signal across the simulation volume.

\subsection{Collective Spectrum of HMXBs}
\label{subsec:collective_spec}
The average collective spectrum of HMXBs in galaxies in the nearby Universe is
\begin{equation}
    L_E = L_0 \times \left(\frac{E}{\rm keV}\right)^{-1.08}\left(\frac{\dot M_*}{\mathrm{M_\odot\,yr^{-1}}}\right)\,\mathrm{erg\,s^{-1}\,keV^{-1}},
    \label{eq:L_-1.08}
\end{equation}
where $L_0 = 1.31 \times 10^{39}$ for local galaxies \citep{2017MNRAS.468.2249S}. We consider three models for the high redshift galaxies:
\begin{enumerate}
\item \textbf{Un-enhanced X-ray emission:} In this scenario, the SFR is $13.8\,\mathrm{M_\odot\,yr^{-1}}$, corresponding to a galaxy in a halo of mass $10^{11}\,\mathrm{M_\odot}$ at $z=12$. Based on this SFR, the photon emission rate of the stellar component, which is represented by a black body radiation source at temperature $10^5\,\rm{K}$, is $S = 3.5\times10^{54}\,\rm{ph\,s^{-1}}$. Eq.~(\ref{eq:L_-1.08}) is adopted for the luminosity of the HMXBs in the galaxy.
    \item \textbf{Enhanced X-ray emission:} This scenario is as in (i), except the luminosity of the HMXBs is enhanced by a factor of ten to account for the low metallicity effect occurring in a high redshift galaxy \citep{2017ApJ...840...39M}, so $L_0 = 1.31 \times 10^{40}$.
    \item \textbf{Enhanced X-ray emission with starburst:} This model considers a situation where the young metal-poor galaxy undergoes a starburst. The SFR is set to be one order of magnitude higher than the previous estimated SFR, so $\rm{SFR} = 138\,\mathrm{M_\odot\,yr^{-1}}$. Consequently, the ionizing photon emission rate of the stellar component and $L_0$ of the HMXBs increase to $S = 3.5\times10^{55}\,\rm{ph\,s^{-1}}$ and $L_0 = 1.31 \times 10^{41}$, respectively. 
\end{enumerate}

\begin{figure*}
\scalebox{0.65}{\includegraphics{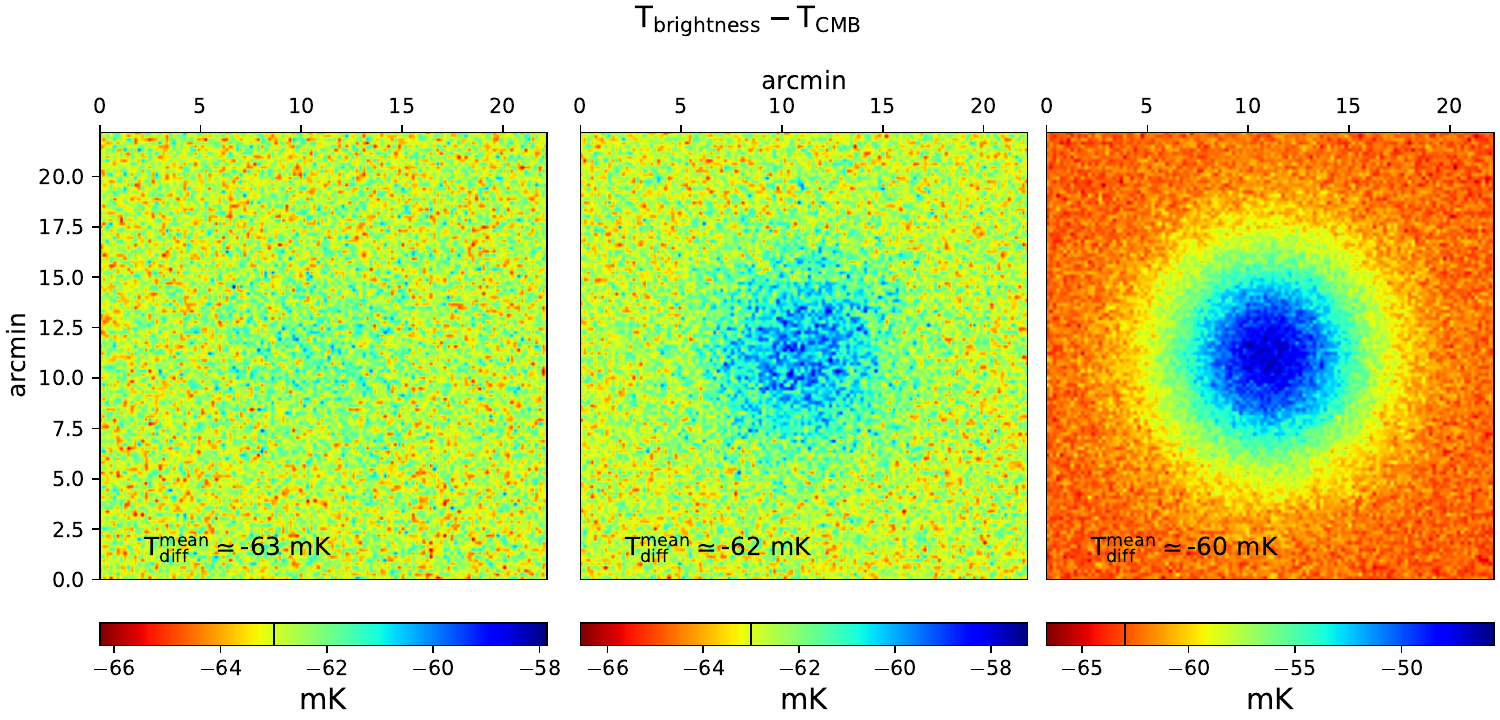}}
%\vspace{-1.5cm}
\caption{The difference between the 21-cm IGM brightness temperature and the CMB as projected on the sky for a galaxy populated by HMXBs with a collective power-law spectrum $L_E\sim E^{-1.08}$, contributed by supersoft, soft and hard sources, at $z=12$. The X-ray luminosity is un-enhanced from the local Universe value in the left panel, and enhanced by an order of magnitude in the central and right panels, as expected for the low metallicity of the galaxy. In the right panel, the star formation rate is additionally boosted by an order of magnitude to mimic a large starburst. All the images have been Gaussian-smoothed and Gaussian noise has been added, with the same smoothing and noise level as in Fig.~\ref{fig:image_21cm_no_burst_BBR_-1.08_R_2.38}. All the background temperature differences are $-63\,\rm{K}$ and the corresponding colour on each panel is indicated by the black line on each colour bar.  In all the images, the mean differential temperature is shown; the CMB wavefront first contacts the causality horizon at $R_0 \simeq 2.38\, \rm{pMpc}$. (The middle panel shows the identical image as the bottom right panel in Fig.~\ref{fig:image_21cm_no_burst_BBR_-1.08_R_2.38}.)
}
\label{fig:image_21cm_I_-1.08_R_2.38}
\end{figure*}

Fig.~\ref{fig:image_21cm_I_-1.08_R_2.38} shows the difference between the brightness temperature of the 21-cm signal and the CMB temperature for these three cases. For the enhanced X-ray emission models, both the galaxy (centre panel) and the starburst case (right-hand panel) exhibit an appreciable 21-cm signature compared with the background differential brightness temperature. For the galaxy model, the differential brightness temperature in the central region differs by about 5~mK from the background value, corresponding to a $\thicksim 5\sigma$ detection level for a noise rms of 1~mK. A surrounding skirt with a difference of about 3~mK is marginally detectable. The starburst model has an even stronger signal, with a $\thicksim 17\sigma$ detection level. By contrast, the 21-cm signature predicted without the X-ray enhancement (left-hand panel) is weak. The maximum difference in the brightness temperature differential (in magnitude) from the background is about $2\,\rm{mK}$. While it is detectable in our simplified calculation, in practice a detection could be challenging due to potential signal contamination from local structures, density variations in the cosmic structure and foreground contamination.

\begin{figure*}
\scalebox{0.65}
{\includegraphics{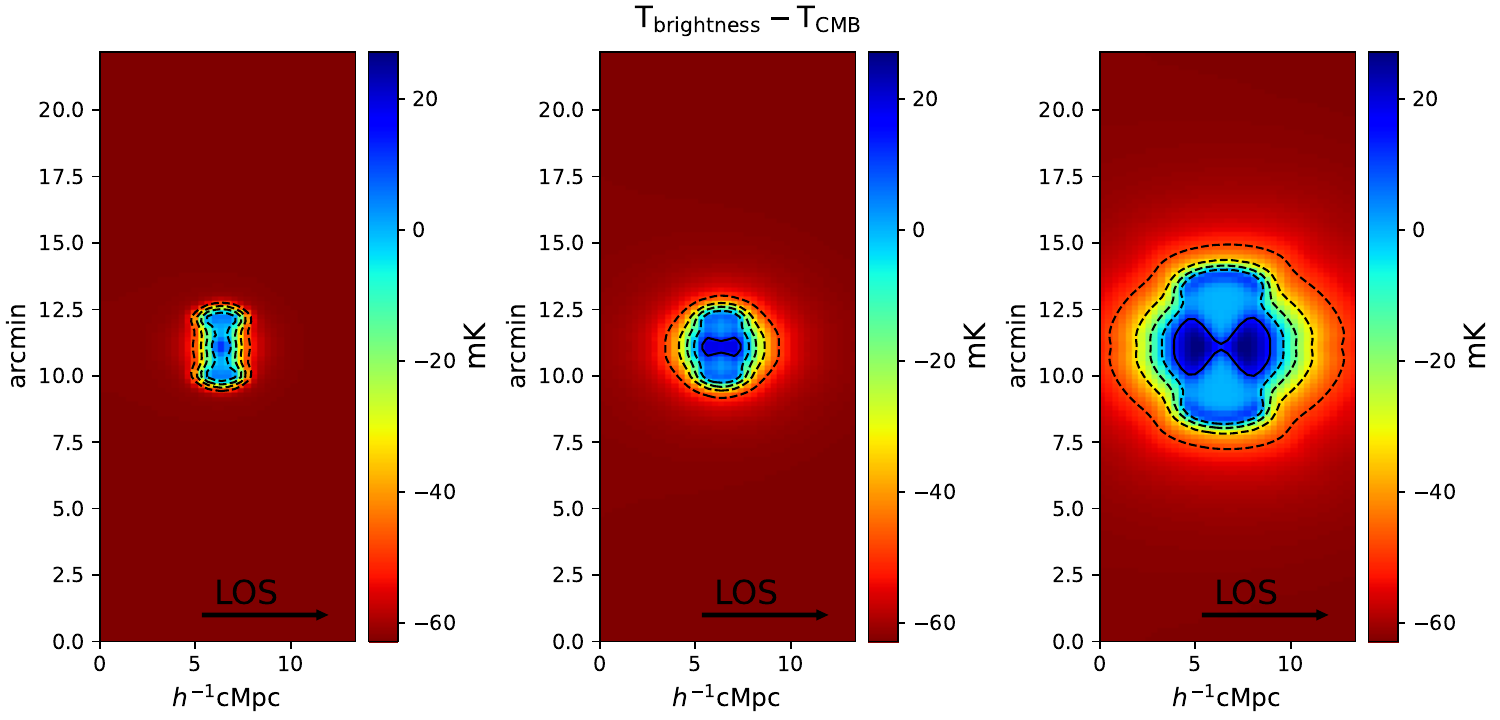}}
%\vspace{-1.5cm}
\caption{Slice plots of the brightness temperature differential without mock observational effects, corresponding to transverse sections through the mid-plane of each respective panel shown in Fig. \ref{fig:image_21cm_I_-1.08_R_2.38}. Panels from left to right are the un-enhanced X-ray luminosity model, the enhanced X-ray luminosity model, and the starburst model, respectively. The slightly asymmetrical profile in the right-hand panel illustrates the time-delay effect, as the X-ray heated region expands: the nearer side (right-hand side, closer to the observer) appears larger in the equal-observing-time frame. The bandwidth corresponding to the width of each panel is $1\,\rm{MHz}$, identical to the projected length used in Fig. \ref{fig:image_21cm_I_-1.08_R_2.38}, and the arrows at the bottom indicate the direction to the observer. The contour values are identical in each panel. From outermost to innermost, the contour levels are $-49$, $-33$, $-18$, $-3$ and $13\,\mathrm{mK}$. }
\label{fig:image_21cm_I_-1.08_cross}
\end{figure*}

Fig.~\ref{fig:image_21cm_I_-1.08_cross} shows the simulated transverse sections through the mid-plane of each panel displayed in Fig. \ref{fig:image_21cm_I_-1.08_R_2.38}, illustrating the intrinsic structure of these signals. Case (ii) (middle panel) has the same photon emission rate from stellar sources as case (i) (left panel), but includes an enhanced X-ray luminosity due to the low metallicity effect, resulting in a larger region decoupled from the background. This broader and brighter area explains why case (ii) produces a stronger signal map (middle panel in Fig.~\ref{fig:image_21cm_I_-1.08_R_2.38}) compared to case (i) (left panel in Fig.~\ref{fig:image_21cm_I_-1.08_R_2.38}). Similarly, case (iii) (right panel), where a starburst is occurring, exhibits the largest and hottest X-ray heated region in all the cases, corresponding to the most distinctive temperature differential from the background (right panel in Fig. \ref{fig:image_21cm_I_-1.08_R_2.38}). Moreover, these plots reveal that non-absorption and emission regions exist around the galaxy. In non-absorption areas, the brightness temperature differential approaches zero where hydrogen is fully ionised, as indicated by Eq. \ref{eq:dTIGM}. On the other hand, in emission regions, hydrogen remains neutral or partially ionised and is additionally heated by the X-ray emitted by HMXBs, causing the brightness temperature to be above the CMB temperature. It is clear that the observed signals significantly depend on the size and geometry of the non-absorption and emission regions. Consequently, degeneracies such as the beam opening angle of the photoionizing radiation and the emission direction of the galaxy may affect the interpretation of observations.

The mix of HMXB spectral types during Cosmic Dawn is unknown. We next consider each spectral type in turn to assess the detectability of galaxies if each spectral type dominates the HMBX population. While the collective spectrum for nearby galaxies adds all three spectral types, we here maintain the same normalisations at $1\,\rm{keV}$ for each spectral type (except in the starburst scenario), so in essence model scenarios in which the other two spectral types are simply absent, as if no stellar evolutionary pathway leads to their formation. (This simple assumption obviates a need to chose some arbitrary means to normalise the different spectra.)  For each spectral type, we consider models for the three cases (i), (ii) and (iii) above.

\subsection{Super-soft power-law spectrum}
\label{subsec:super_soft_spec}
The specific luminosity of HMXBs with a super-soft X-ray power-law spectrum is modelled as 
\begin{equation}
    L_E = L_0 \times \left(\frac{E}{\rm keV}\right)^{-2.72}\left(\frac{\dot M_*}{\mathrm{M_\odot\,yr^{-1}}}\right)\,\mathrm{erg\,s^{-1}\,keV^{-1}},
    \label{eq:L_-2.72}
\end{equation}
where $L_0 = 0.22 \times 10^{39}$ for local galaxies \citep{2017MNRAS.468.2249S}. We continue to adopt a SFR $13.8\,\mathrm{M_\odot\,yr^{-1}}$ for cases (i) and (ii), with a photon emission rate of the stellar component $S = 3.5\times10^{54}\,\rm{ph\,s^{-1}}$. For case (i), we hold $L_0 = 0.22 \times 10^{39}$; for case (ii), we take $L_0 = 0.22 \times 10^{40}$. For case (iii), we take $\rm{SFR} = 138\,\mathrm{M_\odot\,yr^{-1}}$, with a photon emission rate of the stellar component $S = 3.5\times10^{55}\,\rm{ph\,s^{-1}}$, and $L_0 = 2.2 \times 10^{40}$.

\begin{figure*}
\scalebox{0.65}{\includegraphics{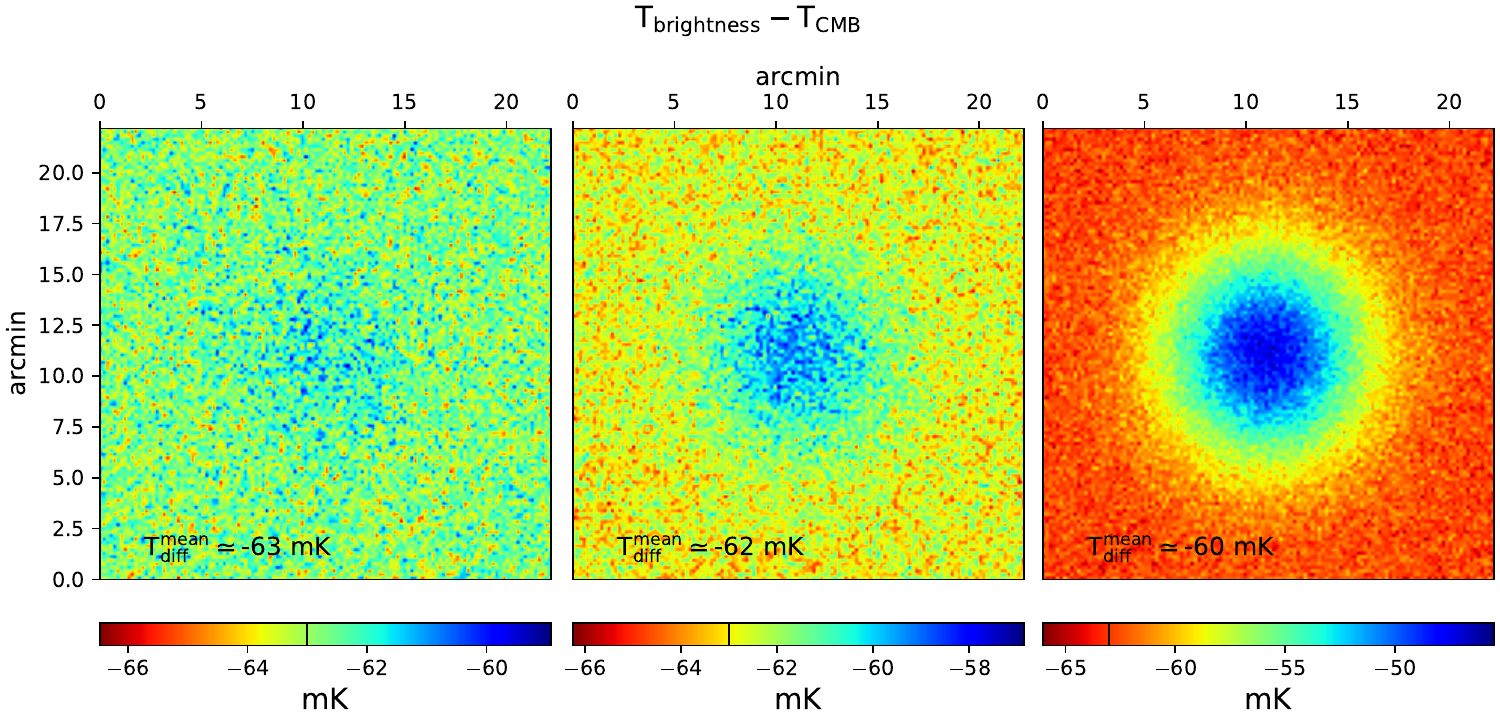}}
%\vspace{-1.5cm}
\caption{The difference between the 21-cm IGM brightness temperature and the CMB as projected on the sky for a galaxy populated by HMXBs with a super-soft X-ray power-law spectrum $L_E\sim E^{-2.72}$ at $z=12$. The three panels correspond to the same cases as in Fig.~\ref{fig:image_21cm_I_-1.08_R_2.38}, with the same smoothing and noise level.
}
\label{fig:image_21cm_I_-2.72_R_2.38}
\end{figure*}

The contrasts in the IGM 21-cm brightness temperature differentials compared with the background are similar to the collective spectrum case, as shown in Fig. \ref{fig:image_21cm_I_-2.72_R_2.38}. The maximum contrast in the brightness temperature differential between the core and the background for the enhanced X-ray luminosity model is again about $5\,\rm{mK}$, and a surrounding skirt with a brightness temperature differential difference from the surroundings of about 2~mK is still discernible. The starburst model displays a larger contrast, as for the collective spectrum case, with the core and the surrounding halo differing from the background by about $17\,\rm{mK}$ and $8\,\rm{mK}$, respectively. The differential brightness temperature profile is clearly detectable for a noise rms of $1\,\rm{mK}$. The contrast between the 21-cm brightness temperature differential of the galaxy without the X-ray enhancement and the background value is again hardly detectable.

\subsection{Soft power-law spectrum}
\label{subsec:soft_spec}
The specific luminosity of HMXBs with a soft power-law X-ray spectrum is modelled as 
\begin{equation}
    L_E = L_0 \times \left(\frac{E}{\rm keV}\right)^{-1.37}\left(\frac{\dot M_*}{\mathrm{M_\odot\,yr^{-1}}}\right)\,\mathrm{erg\,s^{-1}\,keV^{-1}},
    \label{eq:L_-1.37}
\end{equation}
where $L_0 = 0.37 \times 10^{39}$ for local galaxies \citep{2017MNRAS.468.2249S}. We continue to adopt a SFR $13.8\,\mathrm{M_\odot\,yr^{-1}}$ for cases (i) and (ii), with a photon emission rate of the stellar component $S = 3.5\times10^{54}\,\rm{ph\,s^{-1}}$. For case (i), we hold $L_0 = 0.37 \times 10^{39}$; for case (ii), we take $L_0 = 0.37 \times 10^{40}$. For case (iii), we take $\rm{SFR} = 138\,\mathrm{M_\odot\,yr^{-1}}$, with an ionizing photon emission rate of the stellar component $S = 3.5\times10^{55}\,\rm{ph\,s^{-1}}$, and $L_0 = 3.7 \times 10^{40}$.

\begin{figure*}
\scalebox{0.65}{\includegraphics{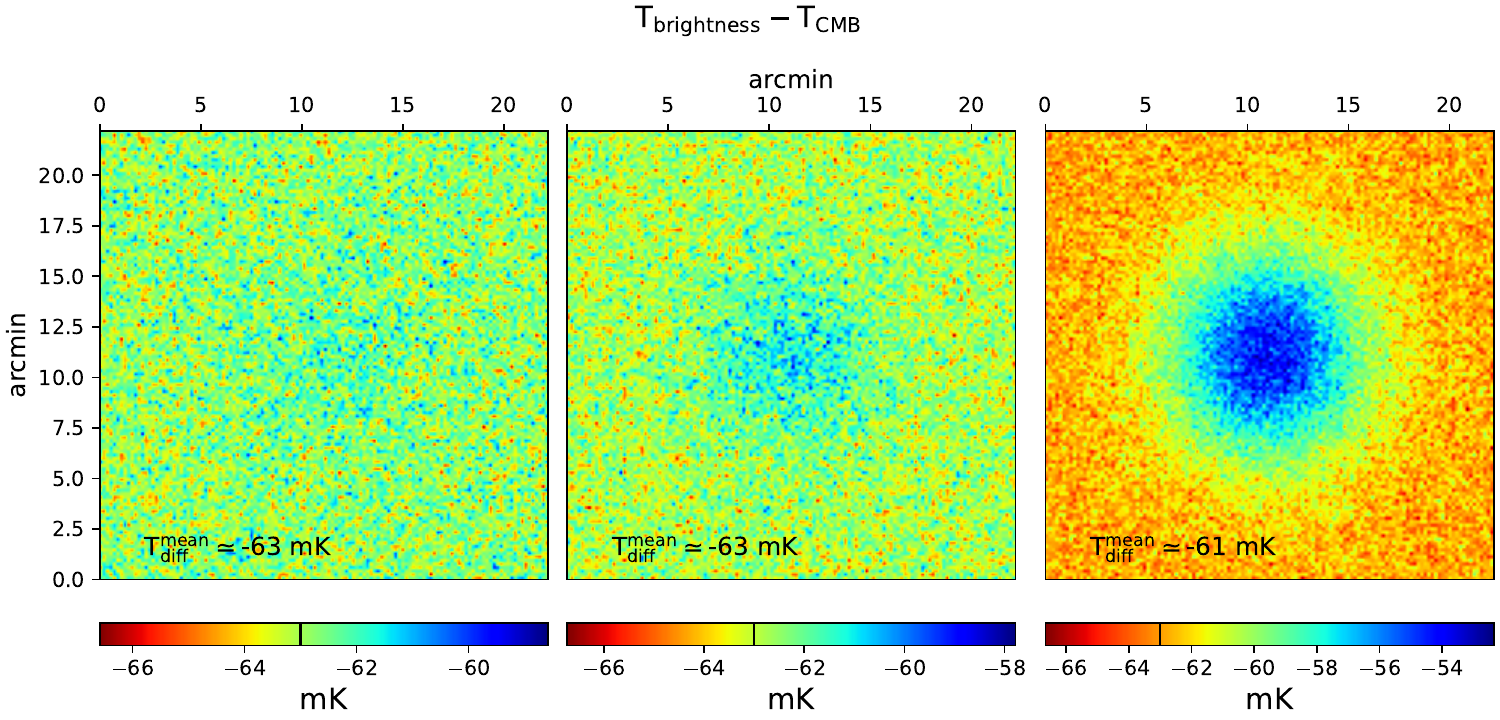}}
%\vspace{-1.5cm}
\caption{The difference between the 21-cm IGM brightness temperature and the CMB as projected on the sky for HMXBs with a soft X-ray power-law spectrum $L_E\sim E^{-1.37}$ at $z=12$, as in Fig.~\ref{fig:image_21cm_I_-1.08_R_2.38}. 
}
\label{fig:image_21cm_I_-1.37_R_2.38}
\end{figure*}

Fig.~\ref{fig:image_21cm_I_-1.37_R_2.38} shows the difference between the brightness temperature of the 21-cm signal and the CMB temperature for these three scenarios. All three cases show considerably diminished signals. The models with the standard star formation rate and un-enhanced and enhanced X-ray luminosities are marginally detectable. The 21-cm signature predicted by the galaxy without the X-ray enhancement has a maximum difference in the brightness temperature differential (in magnitude) from the background of about $2\,\rm{mK}$, but the core is highly mottled:\ the system is hardly detectable. The contrast between the core and background for the case with enhanced X-ray emission shows a maximum difference of about $5\,\rm{mK}$, although the signal in the core is much noisier, with the skirt of lower contrast lost.
%While the core is detectable in our simplified calculation, in practice a detection could be challenging due to potential signal contamination from local structures, density variations in the cosmic structure and foreground contamination.
The starburst model continues to show a more pronounced signal, with a core detectability of $\thicksim 9 \sigma$ for a noise rms of $1\,\rm{mK}$, and a signal radial gradient still clearly discernible. 

\subsection{Hard power-law spectrum}
\label{subsec:hard_soft_spec}
The specific luminosity of HMXBs with a hard X-ray power-law spectrum is modelled as
\begin{equation}
    L_E = L_0 \times \left(\frac{E}{\rm keV}\right)^{-0.49}\left(\frac{\dot M_*}{\mathrm{M_\odot\,yr^{-1}}}\right)\,\mathrm{erg\,s^{-1}\,keV^{-1}},
    \label{eq:L_-0.49}
\end{equation}
where $L_0 = 0.47 \times 10^{39}$ for local galaxies \citep{2017MNRAS.468.2249S}. We continue to adopt a SFR $13.8\,\mathrm{M_\odot\,yr^{-1}}$ for cases (i) and (ii), with a photon emission rate of the stellar component $S = 3.5\times10^{54}\,\rm{ph\,s^{-1}}$. For case (i), we hold $L_0 = 0.47 \times 10^{39}$; for case (ii), we take $L_0 = 0.47 \times 10^{40}$. For case (iii), we take $\rm{SFR} = 138\,\mathrm{M_\odot\,yr^{-1}}$, with an ionizing photon emission rate of the stellar component $S = 3.5\times10^{55}\,\rm{ph\,s^{-1}}$, and $L_0 = 4.7 \times 10^{40}$.

\begin{figure*}
\scalebox{0.65}{\includegraphics{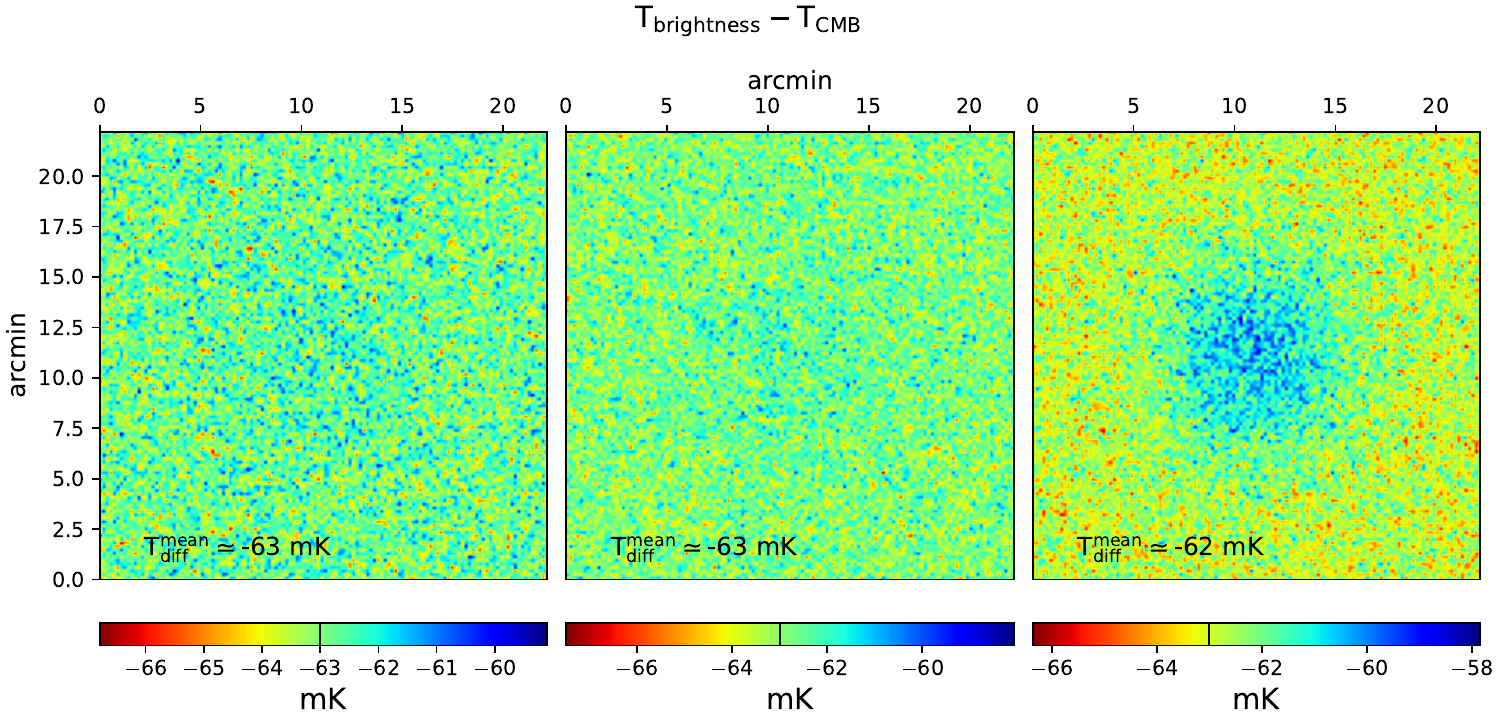}}
%\vspace{-1.5cm}
\caption{The difference between the 21-cm IGM brightness temperature and the CMB as projected on the sky for HMXBs with a hard power-law spectrum $L_E\sim E^{-0.49}$ at $z=12$, as in Fig.~\ref{fig:image_21cm_I_-1.08_R_2.38}.
}
\label{fig:image_21cm_I_-0.49_R_2.38}
\end{figure*}

As shown in Fig.~\ref{fig:image_21cm_I_-0.49_R_2.38}, the IGM 21-cm brightness temperature differentials for the hard HMXB X-ray spectrum are smaller in magnitude than for the other X-ray spectral cases, resulting in lower signal-to-noise ratio detections of the contrast with the background differential temperature. For the galaxy models without and with the X-ray enhancement, the 21-cm signature is undetectable for a noise rms of $1\,\rm{mK}$. By contrast, the starburst model continues to predict a visible 21-cm signature. The difference in the 21-cm brightness temperature differentials between the central core region and the background reaches approximately $4\,\rm{mK}$, demonstrating a $\thicksim 4\sigma$ detection for a noise rms of $1\,\rm{mK}$. While the profile in the signal is no longer pronounced compared with the other spectral cases, a skirt with a contrast of $\sim2$~mK is discernible. The starburst model with the hard HMXB X-ray spectrum offers a detectable signal similar to the galaxy with enhanced X-ray emission and a soft HMXB X-ray spectrum. 

\section{Discussion}
\label{sec:discussion}

\begin{figure*}
\scalebox{0.65}{\includegraphics{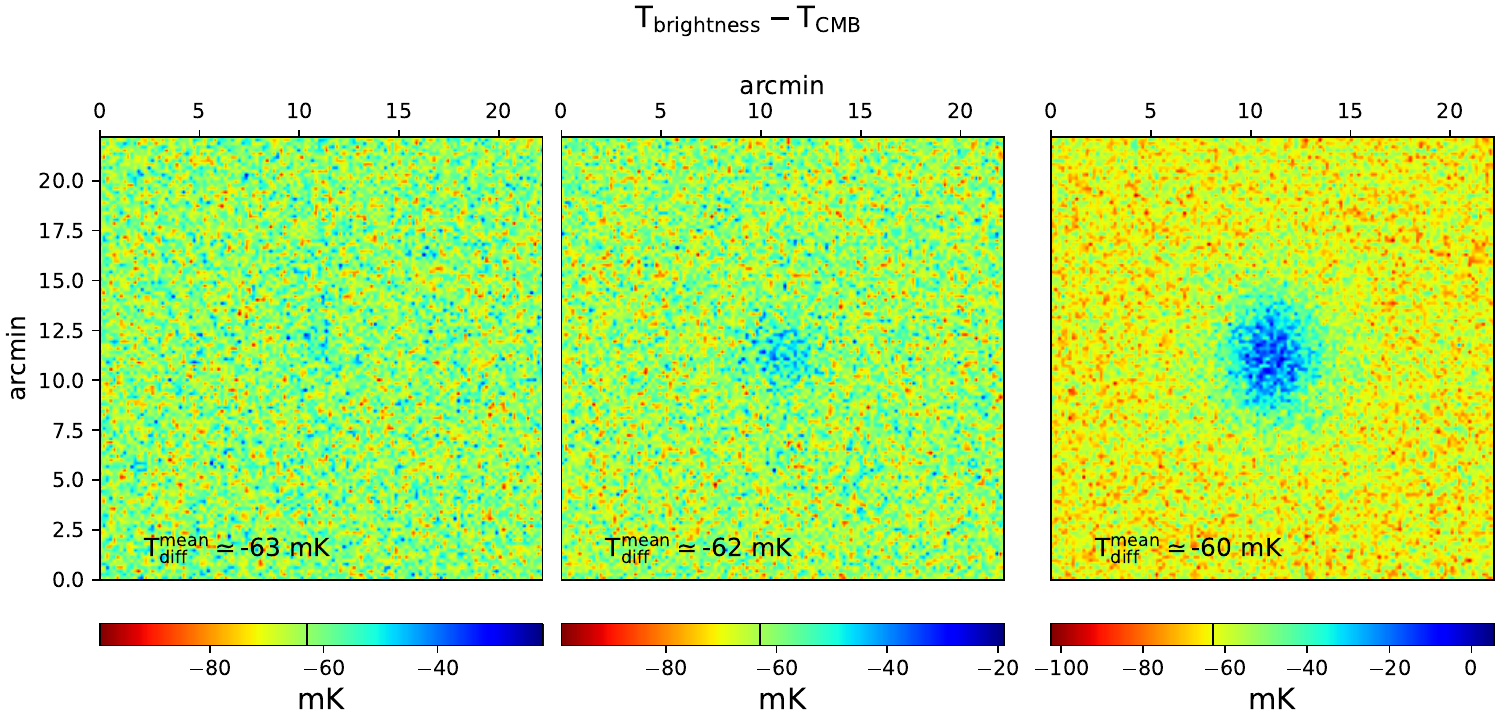}}
%\vspace{-1.5cm}
\caption{The difference between the CMB and 21-cm IGM brightness as projected on the sky for HMXBs with a collective power-law spectrum $L_E\sim E^{-1.08}$. The luminosity profiles of the central galaxies are identical to those modelled in Fig.~\ref{fig:image_21cm_I_-1.08_R_2.38}. Here, the angular resolution in these mock images is enhanced from 7~arcmin to 2~arcmin. The noise level is adjusted to an rms of 10~mK to maintain a 1000~hr integration time.
}
\label{fig:image_21cm_burst_2arcmin_10mK_R_2.38}
\end{figure*}

\begin{figure*}
\scalebox{0.65}{\includegraphics{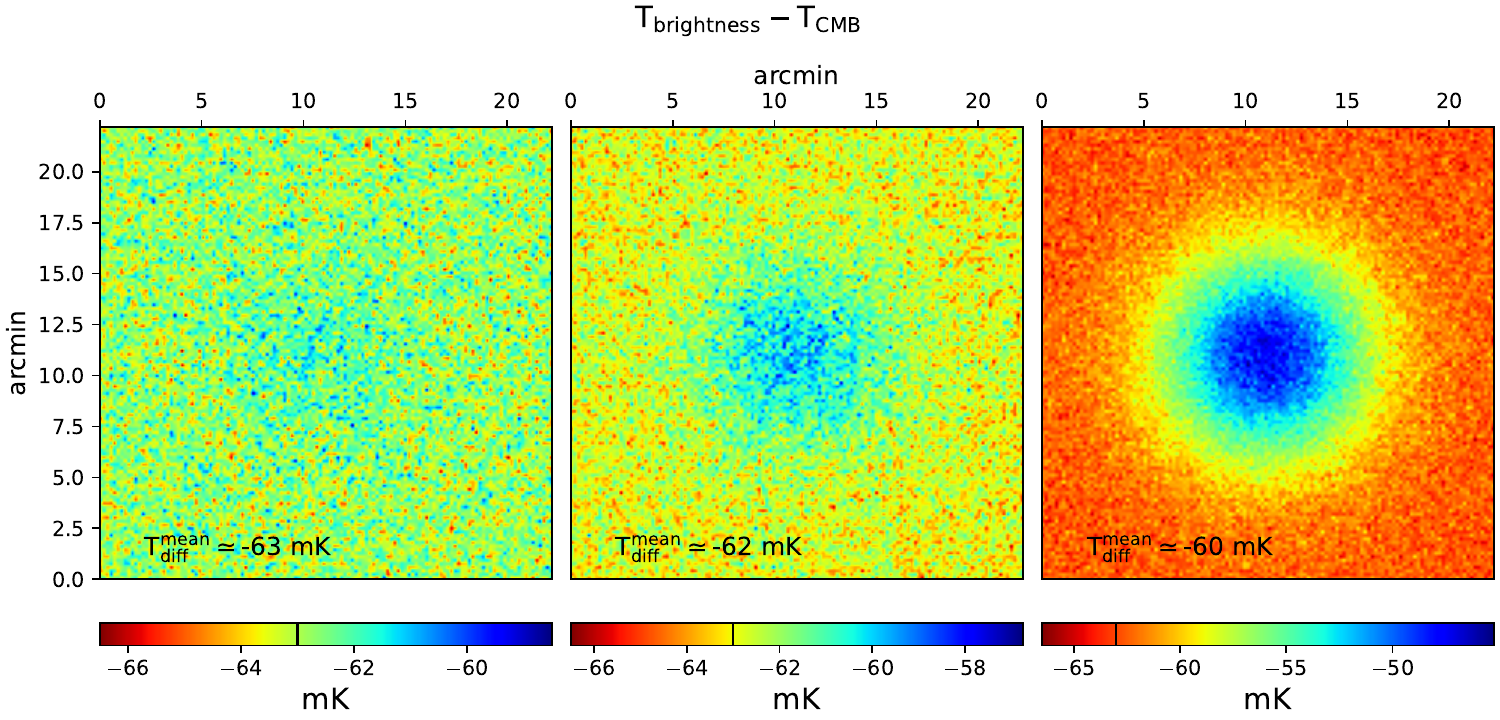}}
%\vspace{-1.5cm}
\caption{The difference between the 21-cm IGM brightness temperature and the CMB as projected on the sky for a galaxy populated by HMXBs with a collective power-law spectrum $L_E\sim E^{-1.08}$ at $z=12$. The three panels correspond to the same cases as in Fig.~\ref{fig:image_21cm_I_-1.08_R_2.38}, with the same smoothing and noise level, except the LOS direction for these three images is parallel to the rotation axis of the galactic disk, and so along the direction of the photoionizing beam.}
\label{fig:image_21cm_I_-1.08_R_2.38_top_down}
\end{figure*}

\begin{figure*}
\scalebox{0.65}{\includegraphics{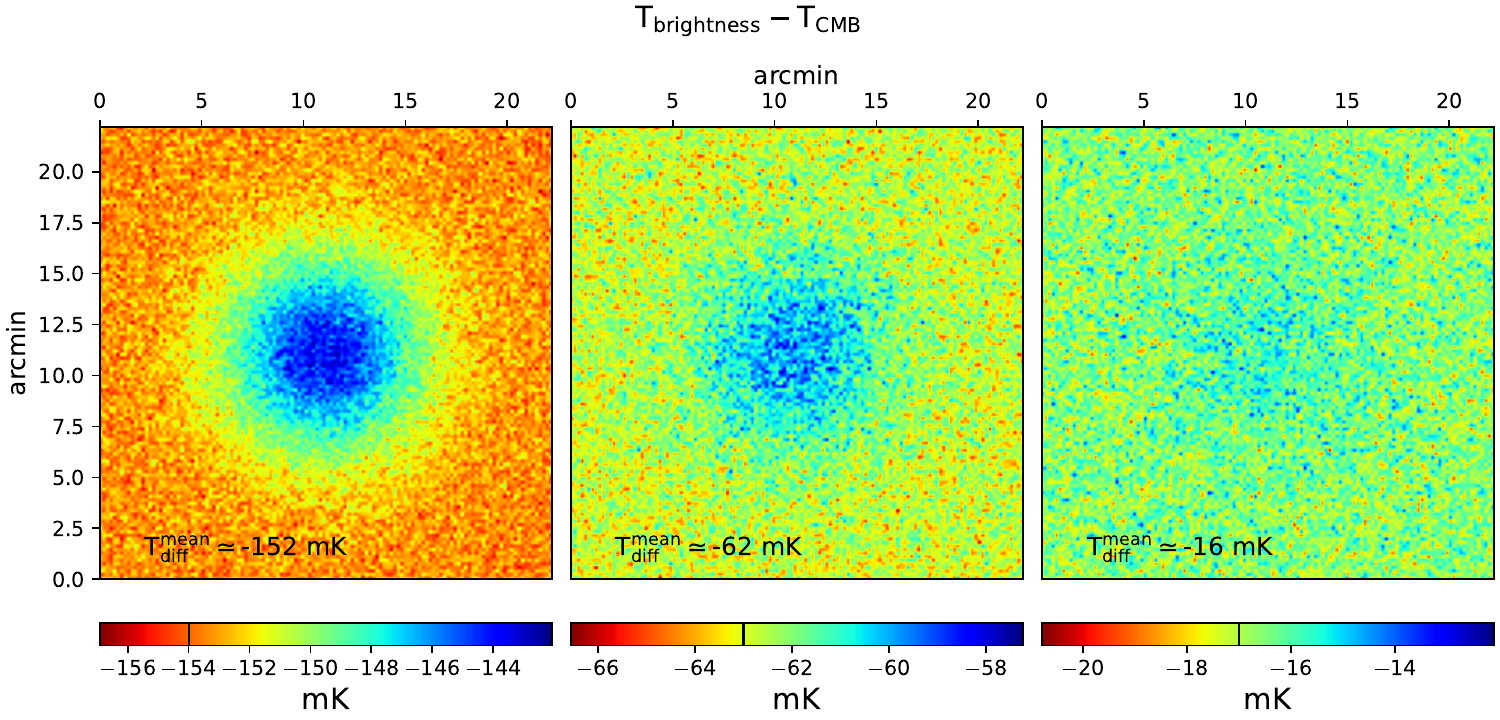}}
%\vspace{-1.5cm}
\caption{The difference between the CMB and 21-cm IGM brightness temperatures as projected on the sky for HMXBs with a collective power-law spectrum $L_E\sim E^{-1.08}$. All the source luminosity profiles are identical, and for a galaxy with enhanced X-ray emission and no starburst (case ii). Here, the background spin temperatures of the intergalactic gas from left to right panels are $6\,\mathrm{K}$, $12\,\mathrm{K}$ and $24\,\mathrm{K}$, respectively. The background values of each differential brightness temperature (indicated by the black line on each colour bar) are about $-154\,\mathrm{K}$, $-63\,\mathrm{K}$ and $-17\,\mathrm{K}$, respectively. Note that the middle panel is the same as the middle panel of Fig.~\ref{fig:image_21cm_I_-1.08_R_2.38}.}
\label{fig:image_21cm_7arcmin_1mK_diff_Ts}
\end{figure*}

The 21-cm signatures of X-ray heating around bright galaxies will be largely isotropic if the X-ray emission from a galaxy is isotropic. It will not necessarily be uniform, however:\ for a sufficiently bright source, the radial profile may be detectable. While we assumed isotropic X-ray emission from the galaxies, as many HMXBs may be in the galactic bulge or halo, we imposed a small opening angle for the ionizing photons from the stellar component, as may result from obscuration by interstellar gas within the galaxy. The photoionizing radiation from the stars is assumed biconical and oriented vertically in the plane of the sky. The biconical nature of the ionizing radiation is not prominent in the images presented above because of the angular resolution and noise level assumed. For high star formation rates, however, SKA2-Low may be able to detect the effect of non-isotropic emission of ionizing photons from a galaxy at higher angular resolution. As shown in Fig.~\ref{fig:image_21cm_burst_2arcmin_10mK_R_2.38}, allowing for a FWHM resolution of 2~arcmin and increased noise level of 10~mK for a 1000~hr integration, using the scaling with angular resolution from \citet{2013ExA....36..235M}, the beaming would be detectable in galaxies undergoing a starburst, with a star formation rate ten times higher than the expectation for a star formation efficiency of one percent. We thus conclude that imaging of bright galaxy targets with SKA2-Low may provide direct tests of the basic assumptions of ionizing photon beam angle and HMXB heating rates of galaxies used to model the predicted 21-cm signature during Cosmic Dawn.

Our viewing angle towards the galaxy may influence the appearance of the observational signature. However, we find that the mock observed images are insensitive to the LOS orientation in our fiducial SKA2-Low configuration. This is because the X-ray heated neutral halo, whose volume significantly exceeds that of the ionised region, dominates the signature (see illustrations in Fig.~\ref{fig:image_21cm_I_-1.08_cross}). Fig.~\ref{fig:image_21cm_I_-1.08_R_2.38_top_down} shows the brightness temperature differentials for the three cases with the collective spectrum ($L_E\sim E^{-1.08}$), but as viewed along the rotation axis of the galactic disks, and so along the photoionizing beam (similar to the top-down perspective in Fig.~\ref{fig:image_21cm_I_-1.08_cross}). The orientation makes little difference:\ the detectability of all three models is comparable to the corresponding cases shown in Fig.~\ref{fig:image_21cm_I_-1.08_R_2.38}.

Since the thermal history of the IGM remains highly uncertain during Cosmic Dawn, the background spin temperature of the IGM may deviate from the fiducial value of $T_{S} = 12\,\rm{K}$ estimated by \citet{2017ApJ...840...39M}. According to Eq.~(\ref{eq:dTIGM}), the observed brightness temperature differential becomes stronger as $T_{S}$ decreases, and vice versa. Fig.~\ref{fig:image_21cm_7arcmin_1mK_diff_Ts} illustrates the 21-cm brightness temperature maps in a 1000~hr integration for three different background IGM spin temperatures, $6\,\rm{K}$, $12\,\rm{K}$ and $24\,\rm{K}$, and a non-starburst galaxy with a collective power-law spectrum $L_E\sim E^{-1.08}$ and enhanced X-ray emission (case ii). The strongest contrast between the core region and the background appears for $T_S = 6\,\mathrm{K}$ (left panel). The signal is weaker for $T_{S}=12\,\rm{K}$ (middle panel), and significantly diminished for $T_{S}=24\,\rm{K}$ (right panel), consistent with the expectation from Eq.~(\ref{eq:dTIGM}). We note that the signature will transition from absorption to emission if the background $T_{S}$ exceeds the CMB temperature.

\begin{figure}
\scalebox{0.5}{\includegraphics{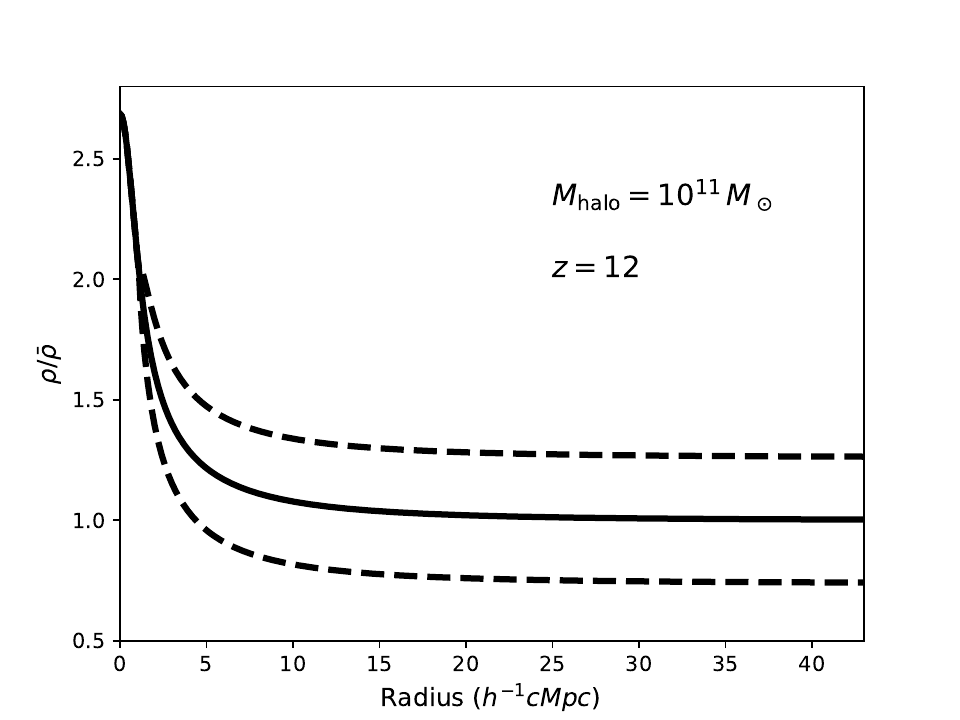}}
%\vspace{-1.5cm}
\caption{The mean matter overdensity around a $10^{11}\,M_\odot$ peak at $z=12$
(solid line) and its rms fluctuations (dashed lines).
}
\label{fig:BBKS_profile}
\end{figure}

Because the source is located in a rare peak in the matter density field, there will be enhanced clustering of matter near it. Ideally, a realistic 3D density profile could be obtained from high-resolution N-body simulations. However, such massive halos are exceedingly rare at $z = 12$ (approximately $14$ per cGpc$^3$ on average), and resolving them at high redshift in $N$-body simulations requires numerous simulations with high mass and spatial resolution, or multiple constrained realisations to sample the most likely fields. To obtain an indication of the role the density profile may play, we here instead place the galaxy at the centre of a density peak with profile given by the mean expected matter profile around a $10^{11}\,M_\odot$ halo at $z=12$ \citep{1986ApJ...304...15B}. As shown in Fig.~\ref{fig:BBKS_profile}, substantial scatter is expected around the peak. Since the profile represents a statistical density enhancement arising from the initial conditions, no adiabatic heating is expected outside of the collapsed halo, and the profile is approximated as spherically symmetric.

\begin{figure*}
\scalebox{0.65}
{\includegraphics{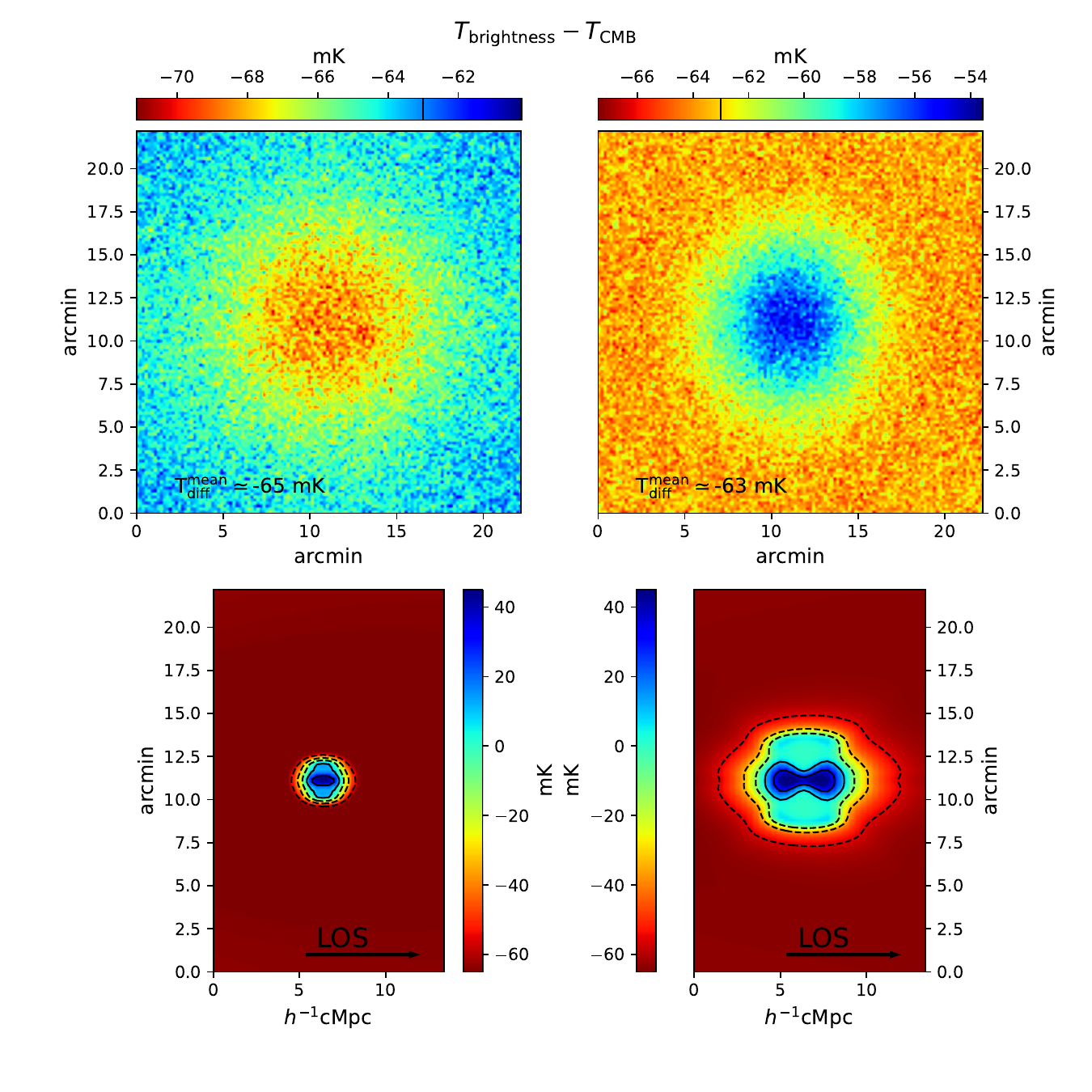}}
\caption{Top panels: The difference between the 21-cm IGM brightness temperature and the CMB projected on the sky for HMXBs with a collective power-law spectrum $L_E\sim E^{-1.08}$ at $z = 12$. The background density follows the angle-averaged density profile (Fig. \ref{fig:BBKS_profile}) given by \citet{1986ApJ...304...15B}. The left panel is for the case (ii), featuring an enhanced X-ray luminosity due to low metallicity, and case (iii), where the low-metallicity galaxy undergoes a starburst, as described in Section \ref{subsec:collective_spec}.  Both images have been Gaussian-smoothed with a FWHM of $7\,{\rm arcmin}$, and Gaussian noise has been added with an rms of $1\,{\rm mK}$. The background temperature differential at mean density is about $-63\,\rm{mK}$ in both images. Bottom panels: The associated transverse sections through the mid-plane of each panel shown above, as in Fig. \ref{fig:image_21cm_I_-1.08_cross}. The contour values are identical in all bottom panels. From outermost to innermost, the contour levels are $-56$, $-35$, $-13$, $8$ and $32\,\mathrm{mK}$. 
}
\label{fig:image_21cm_-1.08_cross_vs_map}
\end{figure*}

As illustrated in Fig.~\ref{fig:image_21cm_-1.08_cross_vs_map}, adopting the angle-averaged density enhancement around the galaxy statistically changes the 21-cm images for case (ii) and case (iii) with the collective spectrum of HMXBs ($L_E\sim E^{-1.08}$). In the starburst scenario (top-right panel), additional gas in the surrounding overdense region enhances the absorption effect, weakening the contrast between the central region and the background temperature to about $10\,\rm{mK}$, corresponding to a $\sim 10\sigma$ detection level for a noise rms of $1\,\rm{mK}$. In contrast, the 21-cm signature for the enhanced X-ray luminosity scenario without starburst (top-left) shows a profound impact from the overdensity profile. Unlike for the uniform density scenario, in which the central signal is brighter than the background signal (the background value is about $-63\,\rm{mK}$), here the central region's brightness temperature is lower than the background. Also, the increased density near the source contracts the \HII\ region around the non-starburst galaxy to $0.2-0.3$~pMpc. The \HII\ region is no longer well-resolved in the simulation, but increasing the spatial resolution to resolve it does not much change the overall signature. The associated transverse section (bottom-left panel) indicates that the X-ray warmed region has contracted compared with the uniform density case, within the chosen bandwidth of $1\,\rm{MHz}$. This suggests the surrounding neutral gas dominates the observed signal, requiring a narrower bandwidth to detect the X-ray heating signature. We checked the signals for lower and higher background IGM temperatures of 6 and 24~K, and the absorption near the source also dominates in these settings. We note that our analysis employs a statistical mean density profile; actual density profiles around real halos will exhibit anisotropic fluctuations, including overdense and underdense regions. Moreover, sources clustering near the galaxy will contribute to the local hydrogen ionization and X-ray heating rates, diminishing the absorption strength. The actual signal for the case without a starburst will be intermediate between the left- and right-hand panels. We conclude that the strength of the signal will depend on the details of the immediate surroundings of the source.

In addition to X-rays from HMXBs, nearby disk galaxies show a comparable amount of diffuse soft X-ray emission believed to originate from the supernovae-heated interstellar medium (ISM) of the galaxy \citep{2012MNRAS.426.1870M}. For ISM densities above $\sim 1\, \mathrm{cm^{-3}}$, metal cooling limits the duration of the hot phase in a supernova remnant, so that the amount of X-ray emission per star formed may be higher for low metallicity galaxies at high redshift \citep{2017MNRAS.471.3632M}, as for HMXBs. The effect of the diffuse X-ray heating on the 21-cm signature may then be comparable to that of the HMXBs, enhancing the signal. The details of the amount of heating, however, are sensitive to the wind modelling, and the amount of internal absorption of the X-rays \citep{2017MNRAS.471.3632M}; accordingly we have concentrated here on the more certain heating estimates for X-rays from HMXBs.

Somewhat smaller mass galaxies than that illustrated here will also show a proximity heating effect in the 21-cm signal, but at lower levels of detectability since the expected star formation rates are lower. Because the abundance of rare haloes rises rapidly with decreasing mass, the greater numbers of the systems may compensate for their lower detectability in terms of their discoverability in an all-sky radio survey, especially as there is a higher chance that some may have unusually high star formation rates, as for starburst galaxies.

\section{Summary and Conclusions}
\label{sec:chapter_4_conclusion}

We have shown the soft X-ray heating from a young bright galaxy during Cosmic Dawn may reduce the 21-cm absorption signature of the still neutral IGM against the CMB in the proximity of the galaxies. We present illustrative cases for which it may be possible to detect a suppression in IGM absorption around rare, bright galaxies during Cosmic Dawn using the SKA. Using a novel fully time-dependent ray-tracing radiative transfer code, we model heating by X-rays from HMXBs for a collective population of hard, soft and super-soft sources based on \emph{Chandra} measurements in the nearby Universe, and including the effects of the absorption of soft X-rays by the IGM outside the expanding photoionized bubble around the galaxy. These different spectra may correspond to different phases in the evolution of HMXBs and to different orientations toward the observer. The mean observed spectrum is only moderately well-established in the present-day Universe, and not at all during Cosmic Dawn. Accordingly, we also show the predicted 21-cm for each spectral class individually. While the 21-cm signature of hard spectral sources may be difficult to detect, except for galaxies with very high rates of star formation, the impact of galaxies dominated by soft or super-soft spectral sources are expected to be more readily discernible. The detection of the proximity heating effect in the 21-cm differential brightness temperature around a galaxy would provide support for an enhanced heating rate by HMXB binaries compared with present-day galaxies, as predicted by models with galactic stars of low metallicity during Cosmic Dawn.

To illustrate the effect, we choose a bright star-forming galaxy in a halo of mass $10^{11}\,M_\odot$ at $z=12$, as the 21-cm signature against the CMB of such a rare object should stand out in contrast to the large-scale 21-cm signature.  
We construct three cases, both with and without the enhanced X-ray emission expected for low-metallicity HMXB stars in high redshift galaxies, and a third with both enhanced X-ray emission and a star formation rate boosted by a factor of ten compared with the nominal expectation for a star formation efficiency of one percent. This corresponds either to a high star formation efficiency (of ten percent) or a burst in star formation.

Except for a starburst galaxy, the required integration times for SKA1-Low to detect the X-ray heating effect from an individual source may be problematically long ($\sim4000$ hrs). We instead focus here on predictions for SKA2-Low. We smooth the models with a 7~arcmin FWHM beam, and add noise with an rms of 1~mK, as expected for a 1000~hr integration using SKA2-Low. For the collective HMXB spectrum, allowing for enhanced X-ray emission by an order of magnitude, as expected for the low metallicity of high redshift galaxies, produces a signal detectable at the $\sim5\sigma$ level, with a radial gradient in the signal discernible, but at a lower level of significance. Boosting the star formation rate by a factor of ten increases the detectability to $\sim17\sigma$, with a clearly detectable radial gradient. By contrast, without enhancing the X-ray emission or the star formation rate, the signal is only marginally detectable at $\sim2\sigma$.

Since the mix of HMXBs spectral types in high redshift galaxies may differ from that at low, we also consider galaxies dominated by each spectral type in turn. This presumes the other spectral types are simply absent, so that the normalisation at 1~keV is correspondingly smaller. The levels of detectability for super-soft spectrum sources are similar to those for the composite spectrum. The larger X-ray luminosity at low energies compensates for the lower normalisation. The models with enhanced X-ray emission continue to be detectable for a galaxy or starburst dominated by soft spectrum HMXBs. By contrast, only the starburst model is detectable when hard spectrum sources dominate.

The models assume isotropic X-ray emission, but photoionization is confined to a biconical beam. For a 7~arcmin beam, the anisotropy in the signal is undetectable for a 1000~hr integration with SKA2-Low. Allowing for a narrower beam of 2~arcmin, with a correspondingly higher noise level of 10~mK, we show that a 1000~hr integration could reveal the asymmetry for the starburst model with a collective HMXB spectrum.

The scenarios considered exhibit some ambiguity in the spectral type of HMXBs that dominate in the interpretation of a weak contrast between the differential 21-cm brightness temperature around a bright galaxy and the background. For a strong contrast, however, the conclusion would be unambiguous:\ no combination of super-soft, soft or hard HMXB X-ray spectra could produce a strong signal without allowing for enhanced HMXB X-ray emission compared with present-day galaxies. A possible exception may be if an ISM heated by supernovae-driven winds produces X-ray emission at a rate much exceeding that of HMXBs in Cosmic Dawn galaxies, contrary to the population of galaxies today.

The signature is also sensitive to the extended density profile expected around a massive halo. For a system with one-percent star formation efficiency, the X-ray heating is unable to overcome the enhanced absorption through the surrounding gas. By contrast, increasing the star formation efficiency by an order of magnitude to produce a starburst results in sufficient X-ray emission to overcome the enhanced density and show the effects of X-ray heating in the 21cm signature. In general, the signature will depend on the structure of the extended gaseous region near the galaxy.

\section*{Acknowledgements}
The authors thank James Bolton and an anonymous referee for helpful comments. KHL acknowledges financial support from the School of Physics and Astronomy, University of Edinburgh. Computations described in this work include some performed
using the \texttt{ENZO} code developed by the Laboratory for Computational Astrophysics at the University of California in San Diego (http://lca.ucsd.edu).
%%%%%%%%%%%%%%%%%%%%%%%%%%%%%%%%%%%%%%%%%%%%%%%%%%
\section*{Data Availability}

No new observational data were generated.

%%%%%%%%%%%%%%%%%%%% REFERENCES %%%%%%%%%%%%%%%%%%

% The best way to enter references is to use BibTeX:

\bibliographystyle{mnras}
\bibliography{ms} % if your bibtex file is called example.bib

\begin{thebibliography}{}
\makeatletter
\relax
\def\mn@urlcharsother{\let\do\@makeother \do\$\do\&\do\#\do\^\do\_\do\%\do\~}
\def\mn@doi{\begingroup\mn@urlcharsother \@ifnextchar [ {\mn@doi@}
  {\mn@doi@[]}}
\def\mn@doi@[#1]#2{\def\@tempa{#1}\ifx\@tempa\@empty \href
  {http://dx.doi.org/#2} {doi:#2}\else \href {http://dx.doi.org/#2} {#1}\fi
  \endgroup}
\def\mn@eprint#1#2{\mn@eprint@#1:#2::\@nil}
\def\mn@eprint@arXiv#1{\href {http://arxiv.org/abs/#1} {{\tt arXiv:#1}}}
\def\mn@eprint@dblp#1{\href {http://dblp.uni-trier.de/rec/bibtex/#1.xml}
  {dblp:#1}}
\def\mn@eprint@#1:#2:#3:#4\@nil{\def\@tempa {#1}\def\@tempb {#2}\def\@tempc
  {#3}\ifx \@tempc \@empty \let \@tempc \@tempb \let \@tempb \@tempa \fi \ifx
  \@tempb \@empty \def\@tempb {arXiv}\fi \@ifundefined
  {mn@eprint@\@tempb}{\@tempb:\@tempc}{\expandafter \expandafter \csname
  mn@eprint@\@tempb\endcsname \expandafter{\@tempc}}}

\bibitem[\protect\citeauthoryear{{Abel} \& {Wandelt}}{{Abel} \&
  {Wandelt}}{2002}]{2002MNRAS.330L..53A}
{Abel} T.,  {Wandelt} B.~D.,  2002, \mn@doi [\mnras]
  {10.1046/j.1365-8711.2002.05206.x}, \href
  {https://ui.adsabs.harvard.edu/abs/2002MNRAS.330L..53A} {330, L53}

\bibitem[\protect\citeauthoryear{{Anninos}, {Zhang}, {Abel}  \&
  {Norman}}{{Anninos} et~al.}{1997}]{1997NewA....2..209A}
{Anninos} P.,  {Zhang} Y.,  {Abel} T.,   {Norman} M.~L.,  1997, \mn@doi [\na]
  {10.1016/S1384-1076(97)00009-2}, \href
  {https://ui.adsabs.harvard.edu/abs/1997NewA....2..209A} {2, 209}

\bibitem[\protect\citeauthoryear{{Aubert} \& {Teyssier}}{{Aubert} \&
  {Teyssier}}{2008}]{2008MNRAS.387..295A}
{Aubert} D.,  {Teyssier} R.,  2008, \mn@doi [\mnras]
  {10.1111/j.1365-2966.2008.13223.x}, \href
  {https://ui.adsabs.harvard.edu/abs/2008MNRAS.387..295A} {387, 295}

\bibitem[\protect\citeauthoryear{{Bardeen}, {Bond}, {Kaiser}  \&
  {Szalay}}{{Bardeen} et~al.}{1986}]{1986ApJ...304...15B}
{Bardeen} J.~M.,  {Bond} J.~R.,  {Kaiser} N.,   {Szalay} A.~S.,  1986, \mn@doi
  [\apj] {10.1086/164143}, \href
  {https://ui.adsabs.harvard.edu/abs/1986ApJ...304...15B} {304, 15}

\bibitem[\protect\citeauthoryear{{Barkana} \& {Loeb}}{{Barkana} \&
  {Loeb}}{2005}]{2005ApJ...626....1B}
{Barkana} R.,  {Loeb} A.,  2005, \mn@doi [\apj] {10.1086/429954}, \href
  {https://ui.adsabs.harvard.edu/abs/2005ApJ...626....1B} {626, 1}

\bibitem[\protect\citeauthoryear{{Bland-Hawthorn} \&
  {Maloney}}{{Bland-Hawthorn} \& {Maloney}}{1999}]{1999ApJ...510L..33B}
{Bland-Hawthorn} J.,  {Maloney} P.~R.,  1999, \mn@doi [\apjl] {10.1086/311797},
  \href {https://ui.adsabs.harvard.edu/abs/1999ApJ...510L..33B} {510, L33}

\bibitem[\protect\citeauthoryear{{Bolton} \& {Haehnelt}}{{Bolton} \&
  {Haehnelt}}{2007}]{2007MNRAS.374..493B}
{Bolton} J.~S.,  {Haehnelt} M.~G.,  2007, \mn@doi [\mnras]
  {10.1111/j.1365-2966.2006.11176.x}, \href
  {https://ui.adsabs.harvard.edu/abs/2007MNRAS.374..493B} {374, 493}

\bibitem[\protect\citeauthoryear{{Bolton}, {Meiksin}  \& {White}}{{Bolton}
  et~al.}{2004}]{2004MNRAS.348L..43B}
{Bolton} J.,  {Meiksin} A.,   {White} M.,  2004, \mn@doi [\mnras]
  {10.1111/j.1365-2966.2004.07567.x}, \href
  {https://ui.adsabs.harvard.edu/abs/2004MNRAS.348L..43B} {348, L43}

\bibitem[\protect\citeauthoryear{{Bond}, {Arnett}  \& {Carr}}{{Bond}
  et~al.}{1984}]{1984ApJ...280..825B}
{Bond} J.~R.,  {Arnett} W.~D.,   {Carr} B.~J.,  1984, \mn@doi [\apj]
  {10.1086/162057}, \href
  {https://ui.adsabs.harvard.edu/abs/1984ApJ...280..825B} {280, 825}

\bibitem[\protect\citeauthoryear{{Braun}, {Bonaldi}, {Bourke}, {Keane}  \&
  {Wagg}}{{Braun} et~al.}{2019}]{2019arXiv191212699B}
{Braun} R.,  {Bonaldi} A.,  {Bourke} T.,  {Keane} E.,   {Wagg} J.,  2019,
  \mn@doi [arXiv e-prints] {10.48550/arXiv.1912.12699}, \href
  {https://ui.adsabs.harvard.edu/abs/2019arXiv191212699B} {p. arXiv:1912.12699}

\bibitem[\protect\citeauthoryear{Bryan et~al.,}{Bryan
  et~al.}{2014}]{Bryan_2014}
Bryan G.~L.,  et~al., 2014, \mn@doi [ApJS] {10.1088/0067-0049/211/2/19}, 211,
  19

\bibitem[\protect\citeauthoryear{{Fialkov}, {Barkana}  \& {Visbal}}{{Fialkov}
  et~al.}{2014}]{2014Natur.506..197F}
{Fialkov} A.,  {Barkana} R.,   {Visbal} E.,  2014, \mn@doi [\nat]
  {10.1038/nature12999}, \href
  {https://ui.adsabs.harvard.edu/abs/2014Natur.506..197F} {506, 197}

\bibitem[\protect\citeauthoryear{{Field}}{{Field}}{1958}]{1958PIRE...46..240F}
{Field} G.~B.,  1958, \mn@doi [Proceedings of the IRE]
  {10.1109/JRPROC.1958.286741}, \href
  {https://ui.adsabs.harvard.edu/abs/1958PIRE...46..240F} {46, 240}

\bibitem[\protect\citeauthoryear{{Fragos}, {Lehmer}, {Naoz}, {Zezas}  \&
  {Basu-Zych}}{{Fragos} et~al.}{2013}]{2013ApJ...776L..31F}
{Fragos} T.,  {Lehmer} B.~D.,  {Naoz} S.,  {Zezas} A.,   {Basu-Zych} A.,  2013,
  \mn@doi [\apjl] {10.1088/2041-8205/776/2/L31}, \href
  {https://ui.adsabs.harvard.edu/abs/2013ApJ...776L..31F} {776, L31}

\bibitem[\protect\citeauthoryear{{Fujita}, {Martin}, {Mac Low}  \&
  {Abel}}{{Fujita} et~al.}{2003}]{2003ApJ...599...50F}
{Fujita} A.,  {Martin} C.~L.,  {Mac Low} M.-M.,   {Abel} T.,  2003, \mn@doi
  [\apj] {10.1086/379276}, \href
  {https://ui.adsabs.harvard.edu/abs/2003ApJ...599...50F} {599, 50}

\bibitem[\protect\citeauthoryear{{Gnedin}, {Kravtsov}  \& {Chen}}{{Gnedin}
  et~al.}{2008}]{2008ApJ...672..765G}
{Gnedin} N.~Y.,  {Kravtsov} A.~V.,   {Chen} H.-W.,  2008, \mn@doi [\apj]
  {10.1086/524007}, \href
  {https://ui.adsabs.harvard.edu/abs/2008ApJ...672..765G} {672, 765}

\bibitem[\protect\citeauthoryear{{Haiman}, {Rees}  \& {Loeb}}{{Haiman}
  et~al.}{1997}]{1997ApJ...484..985H}
{Haiman} Z.,  {Rees} M.~J.,   {Loeb} A.,  1997, \mn@doi [\apj]
  {10.1086/304386}, \href
  {http://ukads.nottingham.ac.uk/abs/1997ApJ...484..985H} {484, 985}

\bibitem[\protect\citeauthoryear{{Heckman}, {Sembach}, {Meurer}, {Leitherer},
  {Calzetti}  \& {Martin}}{{Heckman} et~al.}{2001}]{2001ApJ...558...56H}
{Heckman} T.~M.,  {Sembach} K.~R.,  {Meurer} G.~R.,  {Leitherer} C.,
  {Calzetti} D.,   {Martin} C.~L.,  2001, \mn@doi [\apj] {10.1086/322475},
  \href {https://ui.adsabs.harvard.edu/abs/2001ApJ...558...56H} {558, 56}

\bibitem[\protect\citeauthoryear{{Iliev} et~al.,}{{Iliev}
  et~al.}{2006}]{2006MNRAS.371.1057I}
{Iliev} I.~T.,  et~al., 2006, \mn@doi [\mnras]
  {10.1111/j.1365-2966.2006.10775.x}, \href
  {https://ui.adsabs.harvard.edu/abs/2006MNRAS.371.1057I} {371, 1057}

\bibitem[\protect\citeauthoryear{{Izotov}, {Worseck}, {Schaerer}, {Guseva},
  {Thuan}, {Fricke}  \& {Orlitov{\'a}}}{{Izotov}
  et~al.}{2018}]{2018MNRAS.478.4851I}
{Izotov} Y.~I.,  {Worseck} G.,  {Schaerer} D.,  {Guseva} N.~G.,  {Thuan} T.~X.,
   {Fricke} Verhamme A.,   {Orlitov{\'a}} I.,  2018, \mn@doi [\mnras]
  {10.1093/mnras/sty1378}, \href
  {https://ui.adsabs.harvard.edu/abs/2018MNRAS.478.4851I} {478, 4851}

\bibitem[\protect\citeauthoryear{{Leong}, {Meiksin}, {Lai}  \& {To}}{{Leong}
  et~al.}{2023}]{2023MNRAS.519.5743L}
{Leong} K.-H.,  {Meiksin} A.,  {Lai} A.,   {To} K.~H.,  2023, \mn@doi [\mnras]
  {10.1093/mnras/stac3828}, \href
  {https://ui.adsabs.harvard.edu/abs/2023MNRAS.519.5743L} {519, 5743}

\bibitem[\protect\citeauthoryear{{Lotz}}{{Lotz}}{1967}]{1967ApJS...14..207L}
{Lotz} W.,  1967, \mn@doi [\apjs] {10.1086/190154}, \href
  {https://ui.adsabs.harvard.edu/abs/1967ApJS...14..207L} {14, 207}

\bibitem[\protect\citeauthoryear{{Madau} \& {Fragos}}{{Madau} \&
  {Fragos}}{2017}]{2017ApJ...840...39M}
{Madau} P.,  {Fragos} T.,  2017, \mn@doi [\apj] {10.3847/1538-4357/aa6af9},
  \href {https://ui.adsabs.harvard.edu/abs/2017ApJ...840...39M} {840, 39}

\bibitem[\protect\citeauthoryear{{Madau}, {Meiksin}  \& {Rees}}{{Madau}
  et~al.}{1997}]{1997ApJ...475..429M}
{Madau} P.,  {Meiksin} A.,   {Rees} M.~J.,  1997, \mn@doi [\apj]
  {10.1086/303549}, 475, 429

\bibitem[\protect\citeauthoryear{{Meiksin}}{{Meiksin}}{2011}]{2011MNRAS.417.1480M}
{Meiksin} A.,  2011, \mn@doi [\mnras] {10.1111/j.1365-2966.2011.19362.x}, \href
  {http://adsabs.harvard.edu/abs/2011MNRAS.417.1480M} {417, 1480}

\bibitem[\protect\citeauthoryear{{Meiksin}}{{Meiksin}}{2023}]{2023RNAAS...7...71M}
{Meiksin} A.,  2023, \mn@doi [RNAAS] {10.3847/2515-5172/accbc2}, \href
  {https://ui.adsabs.harvard.edu/abs/2023RNAAS...7...71M} {7, 71}

\bibitem[\protect\citeauthoryear{{Meiksin}, {Khochfar}, {Paardekooper}, {Dalla
  Vecchia}  \& {Kohn}}{{Meiksin} et~al.}{2017}]{2017MNRAS.471.3632M}
{Meiksin} A.,  {Khochfar} S.,  {Paardekooper} J.-P.,  {Dalla Vecchia} C.,
  {Kohn} S.,  2017, \mn@doi [\mnras] {10.1093/mnras/stx1857}, \href
  {https://ui.adsabs.harvard.edu/abs/2017MNRAS.471.3632M} {471, 3632}

\bibitem[\protect\citeauthoryear{{Mellema} et~al.,}{{Mellema}
  et~al.}{2013}]{2013ExA....36..235M}
{Mellema} G.,  et~al., 2013, \mn@doi [Exp. Astron.]
  {10.1007/s10686-013-9334-5}, \href
  {https://ui.adsabs.harvard.edu/abs/2013ExA....36..235M} {36, 235}

\bibitem[\protect\citeauthoryear{{Mineo}, {Gilfanov}  \& {Sunyaev}}{{Mineo}
  et~al.}{2012a}]{2012MNRAS.419.2095M}
{Mineo} S.,  {Gilfanov} M.,   {Sunyaev} R.,  2012a, \mn@doi [\mnras]
  {10.1111/j.1365-2966.2011.19862.x}, \href
  {http://adsabs.harvard.edu/abs/2012MNRAS.419.2095M} {419, 2095}

\bibitem[\protect\citeauthoryear{{Mineo}, {Gilfanov}  \& {Sunyaev}}{{Mineo}
  et~al.}{2012b}]{2012MNRAS.426.1870M}
{Mineo} S.,  {Gilfanov} M.,   {Sunyaev} R.,  2012b, \mn@doi [\mnras]
  {10.1111/j.1365-2966.2012.21831.x}, \href
  {https://ui.adsabs.harvard.edu/abs/2012MNRAS.426.1870M} {426, 1870}

\bibitem[\protect\citeauthoryear{{Norman} \& {Ikeuchi}}{{Norman} \&
  {Ikeuchi}}{1989}]{1989ApJ...345..372N}
{Norman} C.~A.,  {Ikeuchi} S.,  1989, \mn@doi [\apj] {10.1086/167912}, \href
  {https://ui.adsabs.harvard.edu/abs/1989ApJ...345..372N} {345, 372}

\bibitem[\protect\citeauthoryear{{Planck Collaboration}}{{Planck
  Collaboration}}{2018}]{2018arXiv180706209P}
{Planck Collaboration} 2018, preprint, \href
  {http://adsabs.harvard.edu/abs/2018arXiv180706209P} {} (\mn@eprint {arXiv}
  {1807.06209})

\bibitem[\protect\citeauthoryear{{Raiter}, {Schaerer}  \& {Fosbury}}{{Raiter}
  et~al.}{2010}]{2010AA...523A..64R}
{Raiter} A.,  {Schaerer} D.,   {Fosbury} R.~A.~E.,  2010, \mn@doi [\aap]
  {10.1051/0004-6361/201015236}, \href
  {https://ui.adsabs.harvard.edu/abs/2010A&A...523A..64R} {523, A64}

\bibitem[\protect\citeauthoryear{{Reed}, {Bower}, {Frenk}, {Jenkins}  \&
  {Theuns}}{{Reed} et~al.}{2007}]{2007MNRAS.374....2R}
{Reed} D.~S.,  {Bower} R.,  {Frenk} C.~S.,  {Jenkins} A.,   {Theuns} T.,  2007,
  \mn@doi [\mnras] {10.1111/j.1365-2966.2006.11204.x}, \href
  {http://adsabs.harvard.edu/abs/2007MNRAS.374....2R} {374, 2}

\bibitem[\protect\citeauthoryear{{Reynolds}}{{Reynolds}}{1989}]{1989ApJ...339L..29R}
{Reynolds} R.~J.,  1989, \mn@doi [\apjl] {10.1086/185412}, \href
  {https://ui.adsabs.harvard.edu/abs/1989ApJ...339L..29R} {339, L29}

\bibitem[\protect\citeauthoryear{{Robertson} et~al.,}{{Robertson}
  et~al.}{2023}]{2023NatAs...7..611R}
{Robertson} B.~E.,  et~al., 2023, \mn@doi [NatAst]
  {10.1038/s41550-023-01921-1}, \href
  {https://ui.adsabs.harvard.edu/abs/2023NatAs...7..611R} {7, 611}

\bibitem[\protect\citeauthoryear{{Sartorio} et~al.,}{{Sartorio}
  et~al.}{2023}]{2023MNRAS.521.4039S}
{Sartorio} N.~S.,  et~al., 2023, \mn@doi [\mnras] {10.1093/mnras/stad697},
  \href {https://ui.adsabs.harvard.edu/abs/2023MNRAS.521.4039S} {521, 4039}

\bibitem[\protect\citeauthoryear{{Sazonov} \& {Khabibullin}}{{Sazonov} \&
  {Khabibullin}}{2017}]{2017MNRAS.468.2249S}
{Sazonov} S.,  {Khabibullin} I.,  2017, \mn@doi [\mnras]
  {10.1093/mnras/stx626}, \href
  {https://ui.adsabs.harvard.edu/abs/2017MNRAS.468.2249S} {468, 2249}

\bibitem[\protect\citeauthoryear{{Shapley}, {Steidel}, {Pettini}, {Adelberger}
  \& {Erb}}{{Shapley} et~al.}{2006}]{2006ApJ...651..688S}
{Shapley} A.~E.,  {Steidel} C.~C.,  {Pettini} M.,  {Adelberger} K.~L.,   {Erb}
  D.~K.,  2006, \mn@doi [\apj] {10.1086/507511}, \href
  {https://ui.adsabs.harvard.edu/abs/2006ApJ...651..688S} {651, 688}

\bibitem[\protect\citeauthoryear{{Shull} \& {van Steenberg}}{{Shull} \& {van
  Steenberg}}{1985}]{1985ApJ...298..268S}
{Shull} J.~M.,  {van Steenberg} M.~E.,  1985, \mn@doi [\apj] {10.1086/163605},
  \href {https://ui.adsabs.harvard.edu/abs/1985ApJ...298..268S} {298, 268}

\bibitem[\protect\citeauthoryear{{Skinner} \& {Ostriker}}{{Skinner} \&
  {Ostriker}}{2013}]{2013ApJS..206...21S}
{Skinner} M.~A.,  {Ostriker} E.~C.,  2013, \mn@doi [\apjs]
  {10.1088/0067-0049/206/2/21}, \href
  {https://ui.adsabs.harvard.edu/abs/2013ApJS..206...21S} {206, 21}

\bibitem[\protect\citeauthoryear{{Smith} et~al.,}{{Smith}
  et~al.}{2017}]{2017MNRAS.466.2217S}
{Smith} B.~D.,  et~al., 2017, \mn@doi [\mnras] {10.1093/mnras/stw3291}, \href
  {https://ui.adsabs.harvard.edu/abs/2017MNRAS.466.2217S} {466, 2217}

\bibitem[\protect\citeauthoryear{{Sokolowski} et~al.,}{{Sokolowski}
  et~al.}{2022}]{2022PASA...39...15S}
{Sokolowski} M.,  et~al., 2022, \mn@doi [\pasa] {10.1017/pasa.2021.63}, \href
  {https://ui.adsabs.harvard.edu/abs/2022PASA...39...15S} {39, e015}

\bibitem[\protect\citeauthoryear{{Tozzi}, {Madau}, {Meiksin}  \&
  {Rees}}{{Tozzi} et~al.}{2000}]{2000ApJ...528..597T}
{Tozzi} P.,  {Madau} P.,  {Meiksin} A.,   {Rees} M.~J.,  2000, \mn@doi [\apj]
  {10.1086/308196}, \href {http://ads.nao.ac.jp/abs/2000ApJ...528..597T} {528,
  597}

\bibitem[\protect\citeauthoryear{{Verner}, {Ferland}, {Korista}  \&
  {Yakovlev}}{{Verner} et~al.}{1996}]{1996ApJ...465..487V}
{Verner} D.~A.,  {Ferland} G.~J.,  {Korista} K.~T.,   {Yakovlev} D.~G.,  1996,
  \mn@doi [\apj] {10.1086/177435}, \href
  {http://ads.nao.ac.jp/abs/1996ApJ...465..487V} {465, 487}

\bibitem[\protect\citeauthoryear{Wise \& Abel}{Wise \& Abel}{2011}]{Wise_2011}
Wise J.~H.,  Abel T.,  2011, \mn@doi [MNRAS]
  {10.1111/j.1365-2966.2011.18646.x}, 414, 3458

\bibitem[\protect\citeauthoryear{{Wouthuysen}}{{Wouthuysen}}{1952}]{1952AJ.....57R..31W}
{Wouthuysen} S.~A.,  1952, \mn@doi [\aj] {10.1086/106661}, \href
  {http://ads.nao.ac.jp/abs/1952AJ.....57R..31W} {57, 31}

\bibitem[\protect\citeauthoryear{{Wu}, {McQuinn}  \& {Eisenstein}}{{Wu}
  et~al.}{2021}]{2021JCAP...02..042W}
{Wu} X.,  {McQuinn} M.,   {Eisenstein} D.,  2021, \mn@doi [\jcap]
  {10.1088/1475-7516/2021/02/042}, \href
  {https://ui.adsabs.harvard.edu/abs/2021JCAP...02..042W} {2021, 042}

\bibitem[\protect\citeauthoryear{{Wyithe}, {Loeb}  \& {Barnes}}{{Wyithe}
  et~al.}{2005}]{2005ApJ...634..715W}
{Wyithe} J. S.~B.,  {Loeb} A.,   {Barnes} D.~G.,  2005, \mn@doi [\apj]
  {10.1086/497160}, \href
  {https://ui.adsabs.harvard.edu/abs/2005ApJ...634..715W} {634, 715}

\bibitem[\protect\citeauthoryear{{Xu}, {Wise}, {Norman}, {Ahn}  \&
  {O'Shea}}{{Xu} et~al.}{2016}]{2016ApJ...833...84X}
{Xu} H.,  {Wise} J.~H.,  {Norman} M.~L.,  {Ahn} K.,   {O'Shea} B.~W.,  2016,
  \mn@doi [\apj] {10.3847/1538-4357/833/1/84}, \href
  {https://ui.adsabs.harvard.edu/abs/2016ApJ...833...84X} {833, 84}

\makeatother
\end{thebibliography}

% Alternatively you could enter them by hand, like this:
% This method is tedious and prone to error if you have lots of references
%\begin{thebibliography}{99}
%\bibitem[\protect\citeauthoryear{Author}{2012}]{Author2012}
%Author A.~N., 2013, Journal of Improbable Astronomy, 1, 1
%\bibitem[\protect\citeauthoryear{Others}{2013}]{Others2013}
%Others S., 2012, Journal of Interesting Stuff, 17, 198
%\end{thebibliography}

%%%%%%%%%%%%%%%%%%%%%%%%%%%%%%%%%%%%%%%%%%%%%%%%%%

%%%%%%%%%%%%%%%%% APPENDICES %%%%%%%%%%%%%%%%%%%%%

\appendix

\section{Overview of \texttt{3DPhRay}}
\label{appendix:3DPhRay}
\texttt{3DPhRay} is a ray-tracing radiative transfer numerical scheme designed to simulate the reionisation of hydrogen and helium around a central source, explicitly requiring the radiation to propagate at the speed of light. As one of the most luminous radiation sources in the Universe, QSOs emit ionising photons at a sufficient rate to maintain the initial expansion speed of their ionisation bubbles at the speed of light. The luminal expansion phase of these ionisation bubbles can persist until the ionisation bubbles reach a radius of up to $3\, \mathrm{pMpc}$ around the QSO \citep{2023MNRAS.519.5743L}.

\texttt{3DPhRay} solves the 3-dimensional time-dependent radiative transfer equation in static space:

\begin{equation}
  \frac{1}{c}\frac{\partial I_{\nu}}{\partial t} + \hat{n} \cdot \nabla I_{\nu} = - \kappa_{\nu} I_{\nu} + j_{\nu},
	\label{eq:reduced_RT}
\end{equation} 
where $\hat{n}$ is a unit vector along the propagation direction. The time-derivative term in Eq. \ref{eq:reduced_RT} corresponds to a finite propagation speed for the radiation, allowing \texttt{3DPhRay} to accurately model the reionisation of the IGM in a QSO's proximity zone where the ionisation bubbles luminally expand and $\kappa_{\nu}$ rapidly evolves. By implementing the propagation speed of the radiation field, \texttt{3DPhRay} is moreover suitable for simulating phenomena which propagate at the speed of light, such as the expanding warm X-ray heated IGM halo around a massive early galaxy during Cosmic Dawn.

The ray-tracing framework of \texttt{3DPhRay} is developed from the revised version of \texttt{ENZO}'s ray-tracing scheme \citep{Bryan_2014, Wise_2011, 2023MNRAS.519.5743L}, with the crucial enhancement of allowing the propagation speed of rays to be a free parameter. \texttt{3DPhRay} stores photon packages across different timesteps unless they are completely absorbed or escape from the simulation box. Additionally, \texttt{3DPhRay} incorporates the adaptive ray-tracing scheme introduced by \citet{2002MNRAS.330L..53A}, which dynamically increases the angular resolution of the radiation field by splitting rays. This enhancement improves computational efficiency by maintaining stable spatial sampling resolution. \texttt{3DPhRay} numerically guarantees the conservation of photons in the design of the photon package scheme, along with a geometric correction applied to rays traversing grid cells and by using a probabilistic method for photon absorption \citep{Wise_2011, 2004MNRAS.348L..43B, 2023MNRAS.519.5743L}. The conservativity of \texttt{3DPhRay} is validated by simulating a Str\"omgren sphere (Appendix \ref{appendix:Ionizedball}) and the reionisation of the IGM by a QSO (Appendix \ref{appendix:QSO_index-0.5}). The calculations of the ionisation states of \HI, \HII, \HeI, \HeII\ and \HeIII\ are outsourced to the widely used chemistry and cooling solver, \texttt{GRACKLE}\footnote{https://grackle.readthedocs.io/} \citep{2017MNRAS.466.2217S}. Through \texttt{GRACKLE}, \texttt{3DPhRay} supports equilibrium and non-equilibrium calculations, although it solves the ionisation states by the non-equilibrium method by default. Lastly, \texttt{3DPhRay} utilises photoionisation cross-sections from \citet{1997NewA....2..209A}\footnote{The HeI  photoionisation cross-section is updated to that of \citet{1996ApJ...465..487V}.}.

%, using the same configuration adopted by \citet{2023MNRAS.519.5743L}
\subsection{Test problem A:\ Str\"omgren sphere}
\label{appendix:Ionizedball}
We conduct a Str\"omgren sphere simulation to test the conservation of photons by \texttt{3DPhRay}. Through the Str\"omgren sphere simulation, we evaluate whether photon packages are correctly absorbed and whether the chemical solver, \texttt{Grackle}, works appropriately in \texttt{3DPhRay}. We adopt the same simulation configuration described in Appendix C1 of \citet{2023MNRAS.519.5743L}, ensuring consistency with prior studies \citep{2006MNRAS.371.1057I, 2023MNRAS.519.5743L}. 
% whether photon packages are correctly tracked in 3-dimensional space 
The key parameters of the simulation are as follows: the simulation boxsize is $6.6\, \rm{kpc}$ resolved with $128^3$ cells. The number density of hydrogen atoms is $n_{\rm {H}} = 10^{-3}\, \rm{cm}^{-3}$ and the initial temperature of the neutral hydrogen gas is $T = 10^4\, \rm{K}$. The monochromatic radiation source emits photons at the rate $\dot N_\gamma = 5 \times 10^{48}\,\mathrm{s^{-1}}$. The adiabatic index of the gas is set to $\gamma = 1.667$. Moreover, to maintain the gas temperature at $T = 10^4\,\rm{K}$ within the ionised region, the energy of the monochromatic photon is set to $23.26\, \rm{eV}$, which is above the hydrogen ionisation threshold of $13.6\,\rm{eV}$. This ensures the energy received from the ionisation offsets the recombination cooling effects. 

\begin{figure}
    \centering \includegraphics[width=1.0\columnwidth]{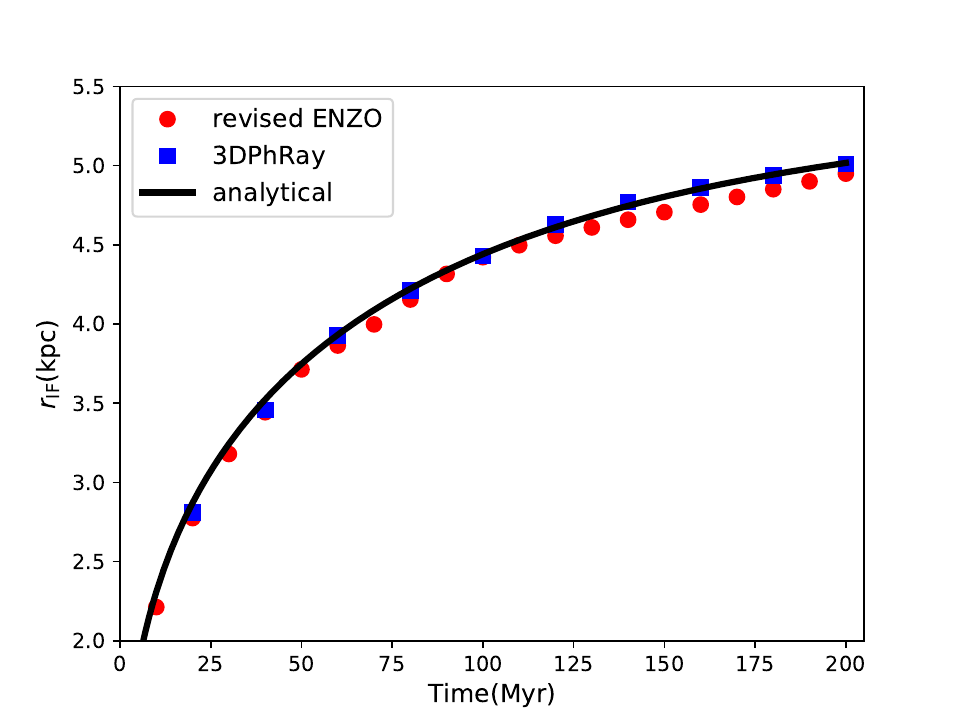}
    \caption{The \HI\ ionisation front of a Str\"omgren sphere. The black solid line shows the analytical solution for the time development of the ionisation front radius $r_{\rm {IF}}$. The red circles show the ionisation radii computed by the revised \texttt{ENZO}, and the blue squares are the results from \texttt{3DPhRay}. The \texttt{3DPhRay} predictions match the analytical solution within a $2\%$ maximum error, validating photon conservation. The configurations used in both simulations are identical to those in the Str\"omgren sphere test shown in \citet{2023MNRAS.519.5743L}.
    }
    \label{fig:HI_front_Stromgren_sphere_3DPhRay}
\end{figure}

Fig. \ref{fig:HI_front_Stromgren_sphere_3DPhRay} illustrates the evolution of the ionisation front radius ($r_{\rm {IF}}$) as predicted by \texttt{3DPhRay}, compared to results from the revised version of \texttt{ENZO} \citep{2023MNRAS.519.5743L} and the analytic solution. Both codes exhibit excellent agreement with the analytic solution. The maximum error between the result predicted by \texttt{3DPhRay} and the analytic solution is about $2\%$, while the revised version of \texttt{ENZO} shows a maximum error of about $4\%$. These results indicate that both simulation codes accurately count photons while tracing rays. The photon conservation capability of \texttt{3DPhRay} demonstrates its reliability.

\subsection{Test problem B: Reionisation by a QSO}
\label{appendix:QSO_index-0.5}
To validate the accuracy of \texttt{3DPhRay} under extreme conditions, we simulate the reionisation processes of hydrogen and helium in the IGM triggered by a QSO. This test allows us to inspect the propagation speed of photon packages in \texttt{3DPhRay} and ensure that photon packages are correctly tracked in 3-dimensional space. The reionisation of the IGM caused by a single QSO involves two phases based on the expansion speed $V_\mathrm{IF}$ of the ionisation fronts:\ a) when the expansion speed exceeds half of the speed of light, $V_{\rm {IF}} > 0.5c$, the reionisation process is in the luminal expansion phase; b) when the expansion speed drops below half of the speed of light, $V_{\rm {IF}} < 0.5c$, the process transitions into the sub-luminal expansion phase. It is worth noting that the expansion speeds for different gas species can differ due to the various ionisation states of the IGM at different redshifts. We refer readers to \citet{2023MNRAS.519.5743L} for more details. 

\subsubsection{Simulation Setup}
The QSO spectrum is modelled as a power-law in frequency, $L_{\nu}=0.56\times10^{31}\,\mathrm{erg s^{-1} Hz^{-1}}(\nu/\nu_L)^{-0.5}$, where $\nu_L$ is the hydrogen photoelectric frequency. The energy range of the spectrum is $13.6\,\rm {eV}$ to $1\, \rm {keV}$. The simulation assumes a uniform background density at redshift $z=6$, so that the hydrogen number density is set to $6.5\times 10^{-5} \rm{cm}^{-3}$. Both hydrogen and helium gases are initially neutral, with an initial temperature of $T = 100 \rm{K}$. The mass-fraction abundances of hydrogen and helium are 0.76 and 0.24, respectively.

\subsubsection{Control Group Simulations}
Due to the absence of an analytic solution (allowing for the temperature evolution), we perform two additional simulations as part of the control group to assess the accuracy of \texttt{3DPhRay}. These simulations use the same configuration as described above but are carried out using a 1D spherically symmetric code, \texttt{PhRay} \citep{2023MNRAS.519.5743L}, and the revised \texttt{ENZO} \citep{2023MNRAS.519.5743L}. The numerical approaches of these two codes differ in some basic assumptions. \texttt{PhRay}\footnote{\texttt{PhRay} is independent of \texttt{3DPhRay}} propagates the radiation field at the speed of light, solving the time-dependent radiative transfer equation (Eq. \ref{eq:reduced_RT}) in 1-dimension with spherical symmetry. The revised \texttt{ENZO} uses a hybrid method introduced by \citet{2023MNRAS.519.5743L}, which transports photon packages at an infinite speed across the entire simulation volume. Although the propagation speed of rays is infinite, during the luminal phase, the hybrid method deletes the photon packages arriving at the transition radius ($R_{t}$), where the expansion speed declines to $0.5c$. During the sub-luminal phase, the hybrid method restricts the photon package travel distance by removing photon packages that exceed their causality horizon. In this test problem, the transition from luminal to sub-luminal expansion occurs at approximately $R_{t} = 3\, \rm{pMpc}$. 

\subsubsection{Spatial and Time Resolutions}
The spatial resolutions of the three simulations differ. The cell size used in the \texttt{PhRay} simulation is about $0.83\, \rm{pkpc}$, ensuring the \HI, \HeI\ and \HeII\ optical depths per zone are less than one even in the neutral region. The \texttt{3DPhRay} simulation adopts a cell size of about $40\, \rm{pkpc}$, balancing numerical precision and computational efficiency. The hybrid simulation uses the biggest cell size, which is about $98\, \rm{pkpc}$. 

The maximum time resolution is restricted by spatial resolution in time-dependent codes like \texttt{PhRay} and \texttt{3DPhRay}. In contrast, time-independent methods, such as the hybrid method of \texttt{ENZO}, are not constrained by their spatial resolution and support an adaptive time step scheme \citep{Wise_2011}. Therefore, to ensure sufficient time resolution, the \texttt{3DPhRay} simulation runs in a higher spatial resolution than the hybrid simulation, providing better accuracy during the luminally expanding phase of the ionisation process.

\subsubsection{Result}
\begin{figure}
    \centering \includegraphics[width=1.0\columnwidth]{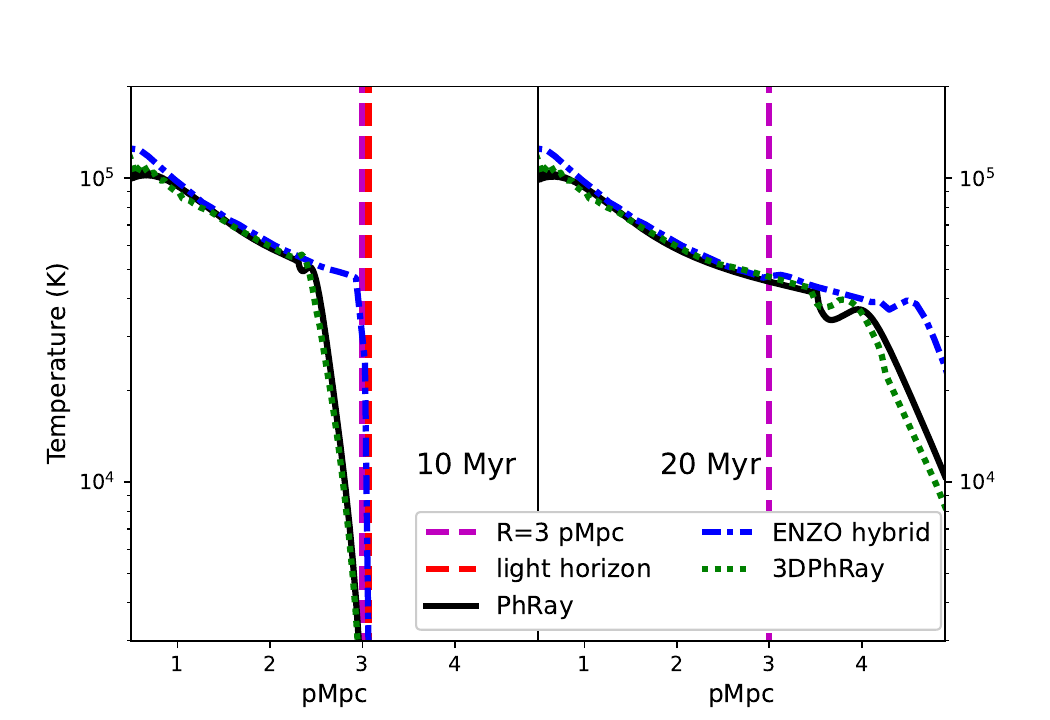}
    \caption{Temperature profiles for reionisation at $z=6$. The vertical red line indicates the causality horizon corresponding to $10\,\rm{Myr}$. The black solid lines, green dotted lines and dot-dashed blue lines are the results of \texttt{PhRay}, \texttt{3DPhRay} and \texttt{ENZO} with the hybrid method, respectively. The temperature profiles of \texttt{PhRay} and \texttt{3DPhRay} match well in both the luminal phase ($10\,\rm{Myr}$) and the sub-luminal phase ($20\, \rm{Myr}$). The temperature front predicted by the \texttt{ENZO} hybrid simulation is somewhat advanced, reflecting the implementation of an infinite propagation speed for the radiation.
    }
    \label{fig:QSO_-0.5_temp}
\end{figure}

\begin{figure}
    \centering \includegraphics[width=1.0\columnwidth]{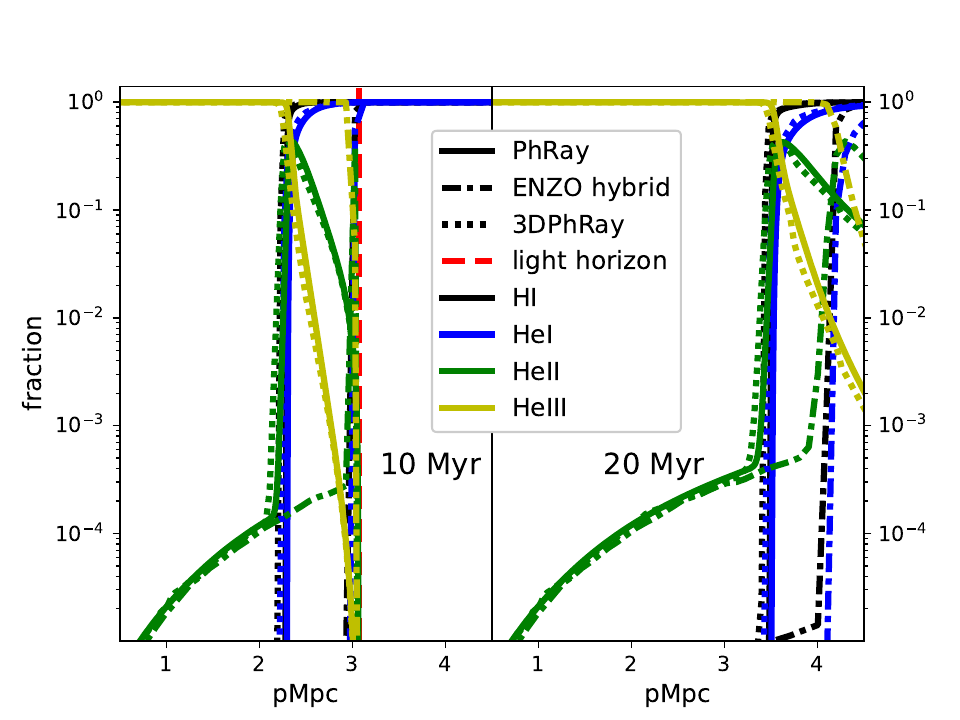}
    \caption{Ionisation profiles for reionisation at $z=6$. The vertical red line indicates the causality horizon corresponding to $10\,\rm{Myr}$. The solid lines, dotted lines and dot-dashed lines are the results of \texttt{PhRay}, \texttt{3DPhRay} and \texttt{ENZO} with the hybrid method, respectively. HI, HeI, HeII and HeIII fractions are coloured as black, blue, green and yellow, respectively. All the ionisation fronts predicted by \texttt{PhRay} and \texttt{3DPhRay} are consistent.  
    }
    \label{fig:fig:QSO_-0.5_fraction}
\end{figure}

Fig. \ref{fig:QSO_-0.5_temp} and Fig. \ref{fig:fig:QSO_-0.5_fraction} show the evolution of the reionisation process in temperature and ionisation fractions over time. In these figures, the \texttt{3DPhRay} simulation results are compared with those from the \texttt{PhRay} and the \texttt{ENZO hybrid} simulations. (The central temperature from \texttt{PhRay} here is slightly below the result in \citet{2023MNRAS.519.5743L} because of a change in the inner boundary condition. In the earlier calculation, a small central region was assumed pre-ionised to avoid the $1/r^2$ flux divergence; this assumption was obviated in the current computations.) All the quantities in the \texttt{3DPhRay} simulations match those in the \texttt{PhRay} simulation, with a slight delay. This latency may be attributed to the difference in spatial resolution between the simulations. The cell size of \texttt{PhRay} is about 48 times smaller than \texttt{3DPhRay}, offering the maximum precision for capturing the luminal ionisation fronts. The \texttt{ENZO hybrid} simulation shows results consistent with those from \texttt{PhRay}, but travelling ahead. This is due to the infinite propagation speed approximation used by \texttt{ENZO}, allowing photon packages to travel instantaneously. 

Overall, this test demonstrates that \texttt{3DPhRay} works appropriately in environments with luminous radiation sources under optically thick conditions. The excellent agreement among these three simulations validates \texttt{3DPhRay}'s capability to model complex reionisation processes.


\section{The role of secondary electron ionisation}
\label{appendix:secondary_ionisation}
Energy losses from collisional ionisation by secondary electrons following X-ray photoionisation diminishes the X-ray heating rate of the still neutral IGM \citep{1997ApJ...475..429M}. To examine the importance of the losses, we compare the temperature profiles of the IGM and the brightness temperature differentials with and without including the secondary electron ionisation effects. We incorporate secondary electron ionisation and the associated X-ray heating suppression into \texttt{3DPhRay} following the power-law limiting approximation of \citet{1985ApJ...298..268S} for ionising photons with energies above $100\,\rm{eV}$. The losses are expected to be local, as the mean free path of ejected electrons with energies between $100-1000$~eV ranges over about $10-100$~pc, using the electron-hydrogen collisional ionisation cross section of \citet{1967ApJS...14..207L}. The gas should be uniform over these scales, as the Jeans length at the mean cosmic mass density $\rho_0$ and adiabatic sound velocity $v_S$ at $z=12$ is $\lambda_J=2\pi v_S/(4\pi G\rho_0)^{1/2}\simeq3(T_\mathrm{IGM}/12\,\mathrm{K})^{1/2}$~kpc.

\begin{figure}
    \centering \includegraphics[width=1.0\columnwidth]{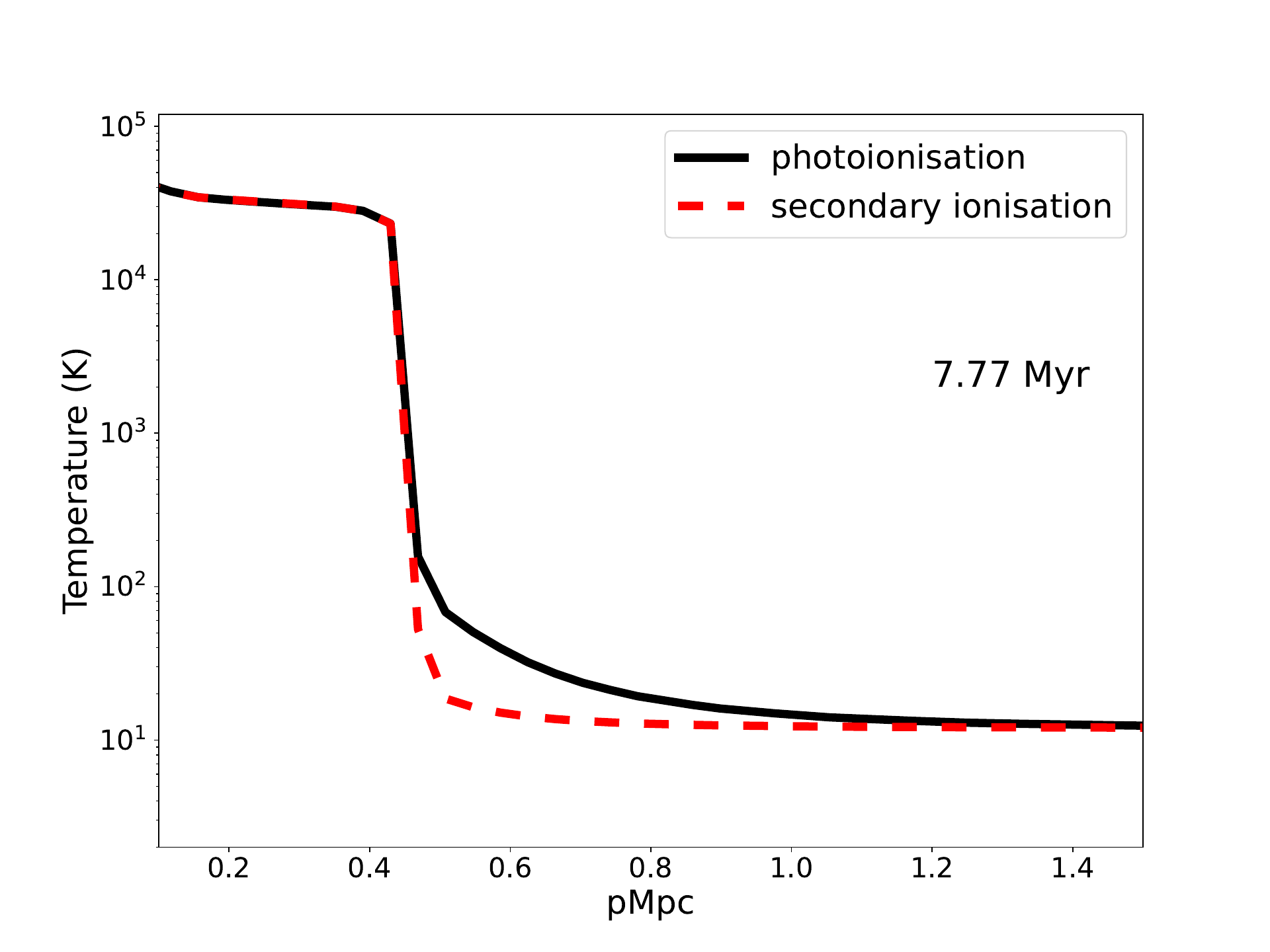}
    \caption{Temperature profiles for IGM around a starburst galaxy with enhanced X-ray emission. The solid black line shows the IGM temperature based on non-equilibrium ionisation calculations that include primary photoionisation but neglect secondary electron ionisation. The dashed red line demonstrates the IGM temperature predicted by non-equilibrium ionisation simulations including primary photoionisation and secondary electron ionisation effects. The suppression in X-ray heating resulting from secondary electron ionisations substantially lowers the temperature of the neutral IGM around the galaxy. (The figure is in the inertial frame of the galaxy at $7.77 \rm{Myr}$.)}
    \label{fig:Temperature_SI_no_SI}
\end{figure}

\begin{figure*}
\scalebox{0.65}
    \centering \includegraphics[width=2.0\columnwidth]{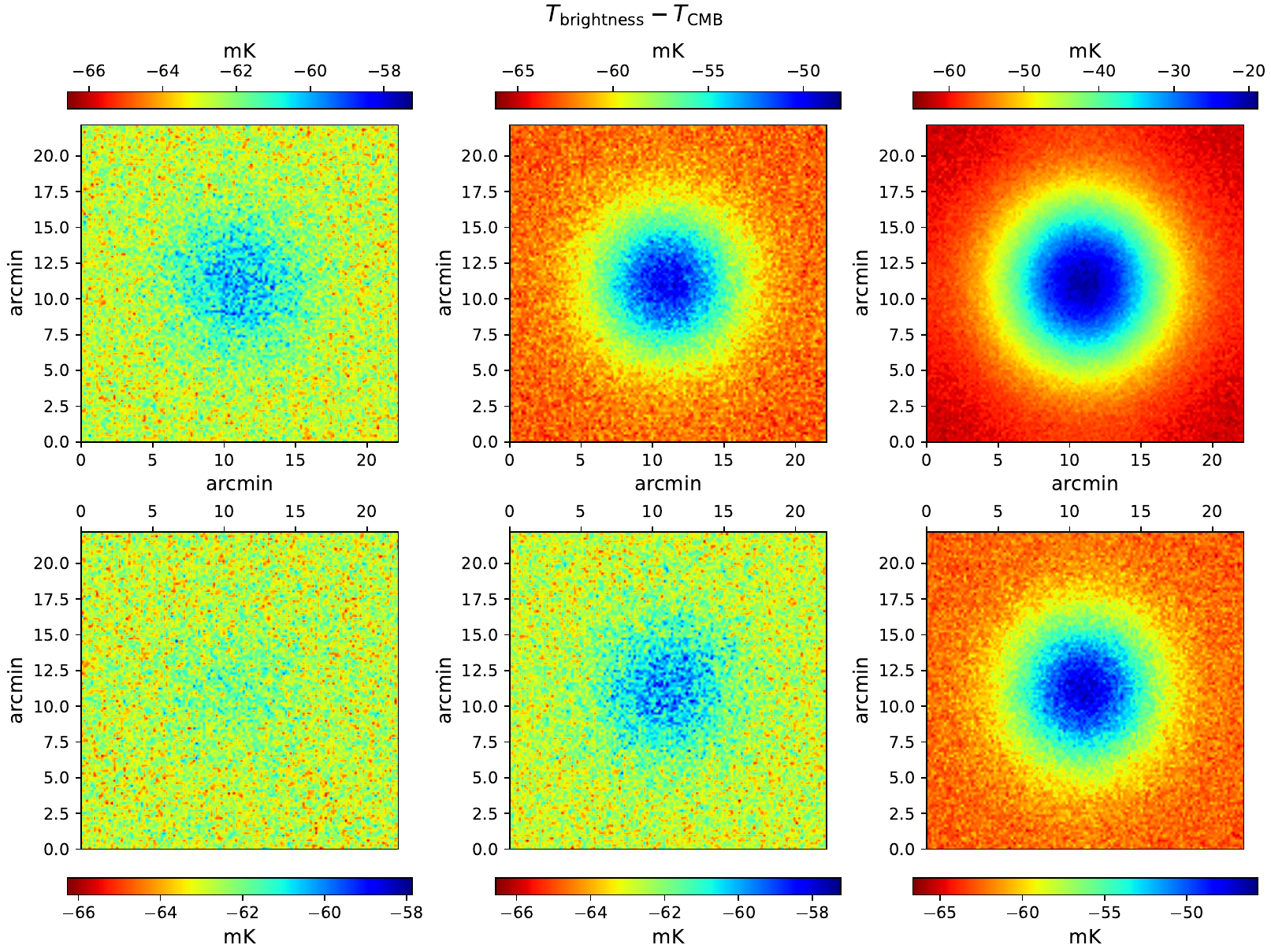}
    \caption{The difference between the 21-cm IGM brightness temperature and CMB temperature as projected on the sky for X-ray heating by HMXBs with a composite power-law spectrum $L_E\sim E^{-1.08}$. The emission models of the central galaxies from the left-hand to the right-hand panels are the un-enhanced X-ray emission model, enhanced X-ray emission model and enhanced X-ray emission with the starburst model (see text). The top row of panels include only photoionisation. The bottom row of panels include additionally secondary electron ionisation. The effects of secondary electron ionisations diminish the 21-cm IGM brightness temperature near the central galaxy.}
    \label{fig:T_map_illustrate_SI}
\end{figure*}

The effect of the suppression in X-ray heating resulting from secondary electron ionisations is illustrated for a starburst galaxy in Fig.~\ref{fig:Temperature_SI_no_SI}. Temperatures are reduced in the surrounding neutral IGM by up to several tens of degrees Kelvin.

The consequences for the 21-cm differential temperature signal $\delta T$ are illustrated in Fig.~\ref{fig:T_map_illustrate_SI} for galaxies with a collective HMXB spectrum (see text). The signal is much suppressed when secondary electron ionisation is included.

\section{DETECTABILITY in SKA1-LOW CONFIGURATION}
\label{appendix:signal_SKA1_low}
For most of the analysis in the paper, we adopt the SKA2-Low configuration to assess the detectability of the 21-cm signatures arising from X-ray heated halos around galaxies during Cosmic Dawn. However, the timeline for SKA2 remains non-definitive, whereas the first phase, SKA1-Low, is expected to be fully commissioned by 2030\footnote{https://www.skao.int/en/647/timeline-science}. In this Appendix, we examine the detectability of the 21-cm signatures using the SKA1-Low AA4 core only model, which comprises $224$ stations\footnote{Based on the SKAO Sensitivity Calculator (\texttt{https://sensitivity-calculator.skao.int})}.

We adopt a sensitivity of $A_\mathrm{eff}/ T_\mathrm{sys}\simeq1$~m$^2$~K$^{-1}$ at $107\,\rm{MHz}$, consistent with the specifications reported in \citet{2019arXiv191212699B} and \citet{2022PASA...39...15S}. To reduce the noise temperature ${T_\mathrm{noise}}$ to an acceptable level, we assume a total integration time of $4000\,\rm{hrs}$. Under a Gaussian point-spread function with FWHM of $7\,\rm{arcmin}$ and a bandwidth of $1\,\rm{MHz}$, the corresponding noise rms is ${T_\mathrm{noise}} \simeq 2.2\,\rm{mK}$, as estimated using eq.~(6) of \citet{2022PASA...39...15S}. Moreover, we further investigate the impact of varying angular resolution and bandwidth on signal detectability by scaling ${T_\mathrm{noise}}$ over different FWHM and bandwidth configurations, aiming to optimise settings for detecting the predicted 21-cm signals. Specifically, we examine the signals produced by HMXBs with a collective power-law spectrum $L_E\sim E^{-1.08}$, using the selected bandwidths and angular resolutions $[0.125, 0.25, 0.5, 1.0]\,\rm{MHz}$ and $[4.0, 6.0, 8.0]\,\rm{arcmin}$, respectively.

\begin{figure*}
\scalebox{0.65}
    \centering \includegraphics[width=2.0\columnwidth]{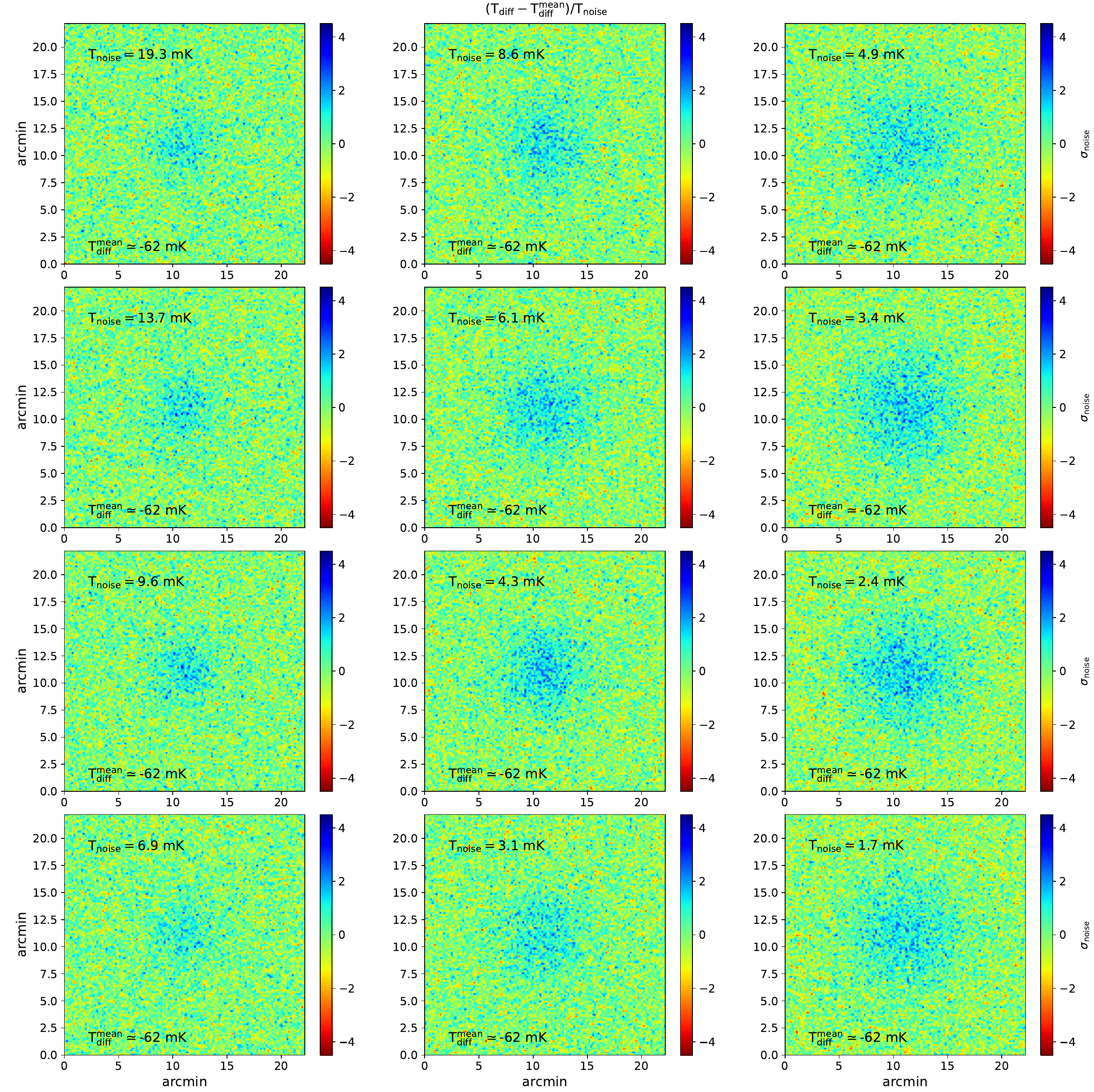}
    \caption{The 21-cm signals, normalised to the noise temperature, for a non-starburst galaxy with a uniform background density at $z = 12$, adopting a collective HMXBs spectrum $L_E \sim E^{-1.08}$. Each panel shows the result for an integration time of $4000\,\mathrm{hrs}$, applying Gaussian smoothing with FWHMs of $4.0$, $6.0$ and $8.0\,\mathrm{arcmin}$ (left to right columns) and bandwidths of $0.125$, $0.25$, $0.5$ and $1.0\,\mathrm{MHz}$ (from top to bottom rows). The image-wide mean 21-cm brightness temperature differential (${T_\mathrm{diff}^\mathrm{mean}}$) and the corresponding noise rms ($T_\mathrm{noise}$) are given in each panel.}
    \label{fig:T_map_SKA1_uniform_no_star_4khr}
\end{figure*}

Fig.~\ref{fig:T_map_SKA1_uniform_no_star_4khr} shows the normalised signal, defined as $({T_\mathrm{diff}} - {T_\mathrm{diff}^\mathrm{mean}}) / {T_\mathrm{noise}}$, where ${T_\mathrm{diff}}$ is the brightness temperature differential, ${T_\mathrm{diff}^\mathrm{mean}}$ is its image-wide mean and ${T_\mathrm{noise}}$ is the noise rms, for a non-starburst galaxy embedded in a uniform background density at $z = 12$. The X-ray luminosity is enhanced allowing for low-metallicity, and the assumed luminosity profile is identical to case (ii) in Sec.~\ref{subsec:collective_spec}. For an angular resolution of $4\,\mathrm{arcmin}$ (left column) across all the selected bandwidths (ascending from top to bottom), the signatures reach about $2\,\sigma$ and are highly mottled, making them difficult to detect in practice. In contrast, for angular resolutions of $6\,\rm{arcmin}$ (middle column) and $8\,\rm{arcmin}$ (right column), the signatures are clearer. Although these signals remain noisy and are limited to $2\,\sigma$ in amplitude, the larger beam sizes enlarge the regions decoupled from the CMB and reduce the overall noise of the image. Among all the cases considered, the signals are most prominent in the $6 - 8\,\rm{arcmin}$ angular resolution range (middle and right columns) combined with bandwidths of $0.25 - 0.5\,\rm{MHz}$ (second and third rows).

\begin{figure*}
\scalebox{0.65}
    \centering \includegraphics[width=2.0\columnwidth]{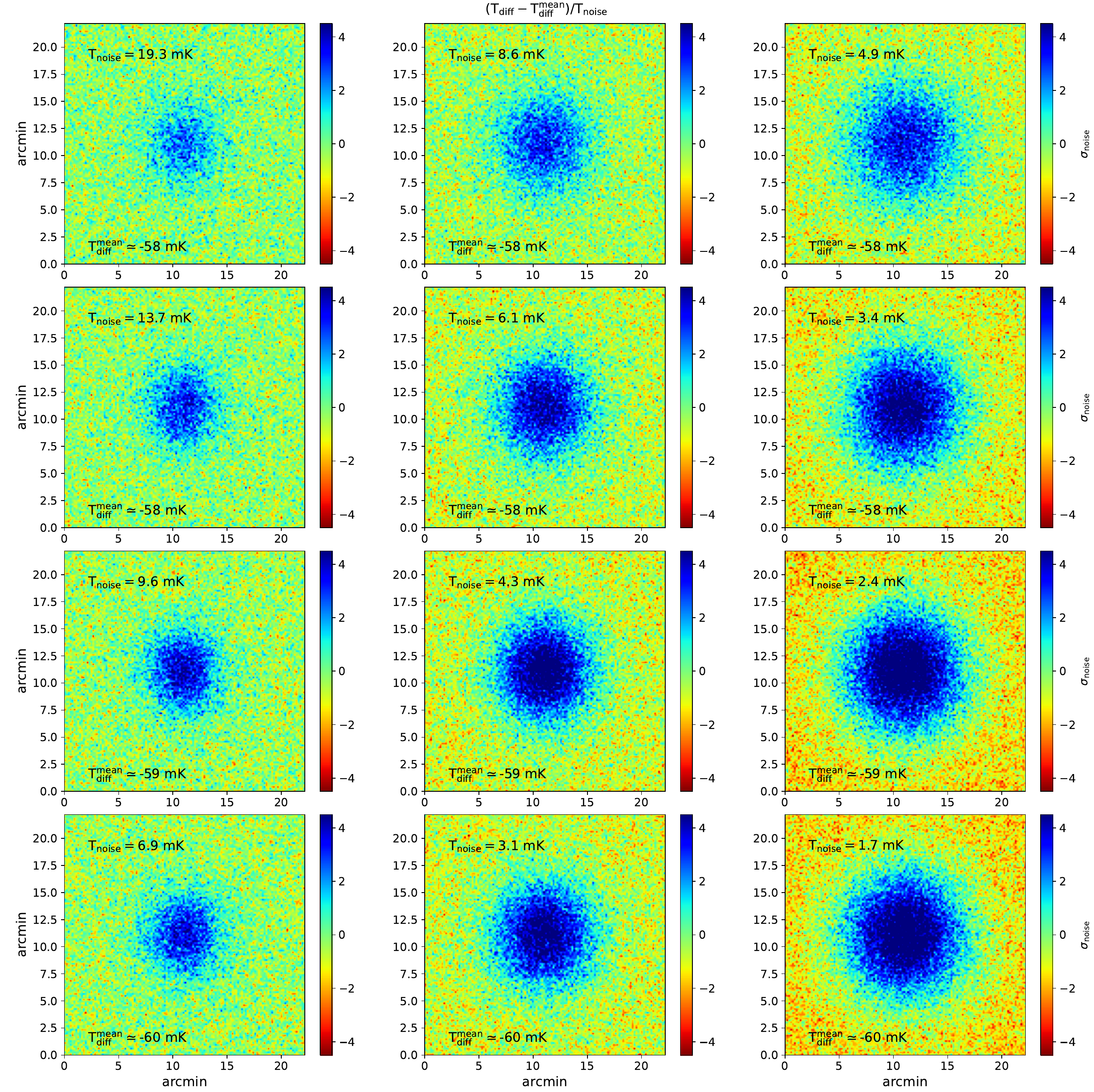}
    \caption{The normalised signals for a starburst galaxy with a uniform background density at $z = 12$, adopting a collective HMXBs spectrum $L_E \sim E^{-1.08}$. All the panels correspond to the same observational setting as in Fig.~\ref{fig:T_map_SKA1_uniform_no_star_4khr}.}
    \label{fig:T_map_SKA1_uniform_star_4khr}
\end{figure*}

\begin{figure*}
\scalebox{0.65}
    \centering \includegraphics[width=2.0\columnwidth]{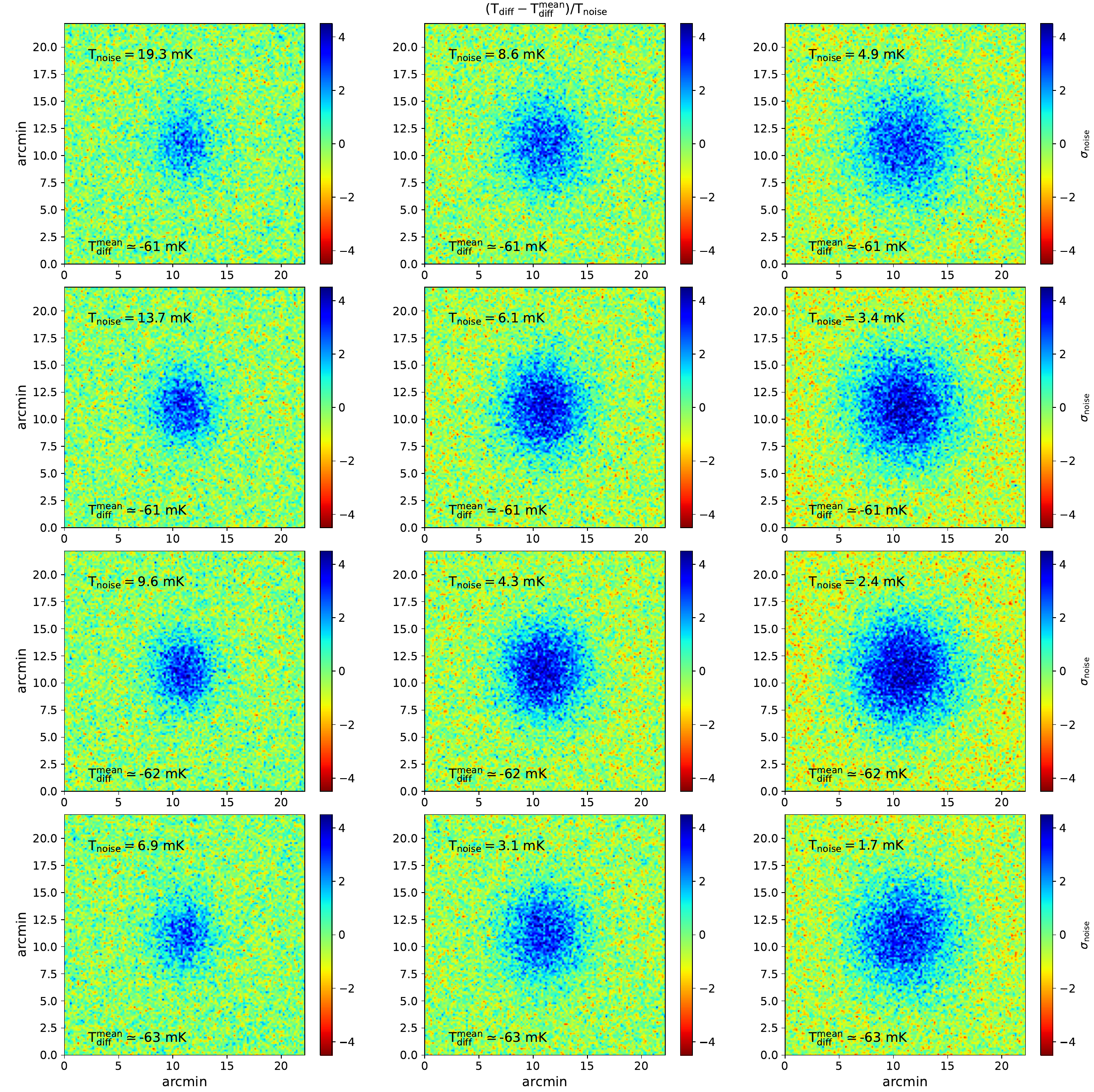}
    \caption{The normalised signals for a starburst galaxy at $z = 12$ with the mean matter overdensity background around a $10^{11}\,M_\odot$ peak, adopting a collective HMXBs spectrum $L_E \sim E^{-1.08}$. All the panels correspond to the same observational setting as in Fig.~\ref{fig:T_map_SKA1_uniform_no_star_4khr}.}
    \label{fig:T_map_SKA1_non_uniform_star_4khr}
\end{figure*}

Figs.~\ref{fig:T_map_SKA1_uniform_star_4khr} and \ref{fig:T_map_SKA1_non_uniform_star_4khr} present the normalised signals for a starburst galaxy with an enhanced X-ray luminosity (corresponding to case (iii) in Sec.~\ref{subsec:collective_spec}) at $z=12$, placed in a uniform density background and the statistically enhanced density around a $10^{11}\,M_\odot$ peak (Fig.~\ref{fig:BBKS_profile}), respectively. For angular resolutions of $6 - 8\,\rm{arcmin}$ (middle and right columns) and bandwidths of $0.25 - 0.5\,\rm{MHz}$ (second and third rows), the amplitudes of the cores reach $3-4\,\sigma$, suggesting this parameter range to investigate in future SKA1-low observations. Furthermore, all the images in both situations exhibit stronger and more extended signatures than those in Fig.~\ref{fig:T_map_SKA1_uniform_no_star_4khr}, confirming the signals produced by star-burst galaxies are more detectable. However, it is important to note that our model assumes idealised observational conditions. In practice, the instrumental designs and foreground contamination potentially reduce the detectability.

%%%%%%%%%%%%%%%%%%%%%%%%%%%%%%%%%%%%%%%%%%%%%%%%%%

% Don't change these lines
\bsp	% typesetting comment
\label{lastpage}
\end{document}